\documentclass{article}

\usepackage[main, final]{neurips_2026_vericode}

\usepackage[utf8]{inputenc} 
\usepackage[T1]{fontenc}    
\usepackage{hyperref}       
\usepackage{url}            
\usepackage{booktabs}       
\usepackage{amsfonts}       
\usepackage{microtype}      
\usepackage{xcolor}         
\usepackage{graphicx}       
\usepackage{longtable}      
\usepackage{listings}       
\usepackage{tikz}           
\usetikzlibrary{shapes.misc}

\usepackage{booktabs}   
\usepackage{multirow}   

\definecolor{lstkw}{HTML}{6B3480}
\definecolor{lstnum}{HTML}{9A9A9A}
\definecolor{lstcomment}{HTML}{7A7A7A}
\definecolor{taxConcurrency}{HTML}{087F8C}
\definecolor{taxLogic}{HTML}{1A6FC4}
\definecolor{taxInput}{HTML}{C8860D}
\definecolor{taxState}{HTML}{7B3F8F}
\definecolor{taxAccess}{HTML}{C44E52}
\newcommand{\tightlist}{%
  \setlength{\topsep}{2pt}\setlength{\partopsep}{0pt}%
  \setlength{\itemsep}{1pt}\setlength{\parsep}{0pt}\setlength{\parskip}{0pt}%
}

\DeclareRobustCommand{\bugtag}[2]{%
  \tikz[baseline=(bugtag.base)]{%
        \node[fill=#1, rounded rectangle, rounded rectangle arc length=180,
          inner xsep=4pt, inner ysep=1.5pt, align=center,
          text height=1.45ex, text depth=.35ex]
      (bugtag) {\color{white}\sffamily\scriptsize\bfseries #2};}}

\title{Spec-Driven PBT: Deriving Semantic Invariants from Code Specifications for Agentic Property-based Testing}

\author{%
  Aditya Shukla \\
  Georgian AI Lab \\
  \And
  Kartik Gupta \\
  Georgian AI Lab \\
  \And
  Ali Alavi \\
  Generative Models Inc. \\
  \And
  Pashootan Vaezipoor \\
  Georgian AI Lab \\
}

\begin{document}

\maketitle

\begin{abstract}
Software is increasingly written by AI agents that autonomously produce an implementation from a user's natural-language description (the specification), a practice known as agentic coding. Code is now produced faster than it can be reviewed, and code written this way has been shown to exhibit its own characteristic failure patterns. Property-based testing (PBT), a long-standing technique in which invariant properties of a program are checked against many generated inputs, has recently been revisited with LLM agents authoring the properties, thereby removing the high-effort derivation of invariants that has long hindered its adoption. Existing agentic PBT methods, however, derive these properties from the implementation alone (Code-driven PBT). Agentic coding presents an opportunity here: a thorough specification, whether a detailed prompt or a full product requirements document (PRD), has become central to the practice, yet its role in testing the resulting code remains underexplored. In this work, we study the value of the specification as a source of properties for testing agent-written code. Our key insight is that the specification, unlike the implementation, records what the code was meant to do, and invariants derived from it are not biased by how a particular implementation happens to do it. We propose Spec-Driven PBT, an agent that first expresses invariants in natural language from the specification and only then inspects the implementation to translate them into concrete tests. Across 25 applications from the ViBench dataset, Spec-Driven PBT finds a novel set of bugs that Code-driven PBT misses, while writing 2.4$\times$ fewer property tests. Though the two approaches surface a roughly equivalent number of bugs overall, there is little overlap: nearly 60\% of the 169 total bugs are found by only one of them. Qualitatively, we find that the bugs unique to Spec-Driven PBT tend to violate documented product workflows whereas those unique to Code-driven PBT instead concentrate on implementation robustness, offering complementary value for bug discovery in agent-written code.
\end{abstract}

\section{Introduction}
\label{sec:intro}
\begin{figure}[t]
  \centering
  \includegraphics[width=0.95\linewidth]{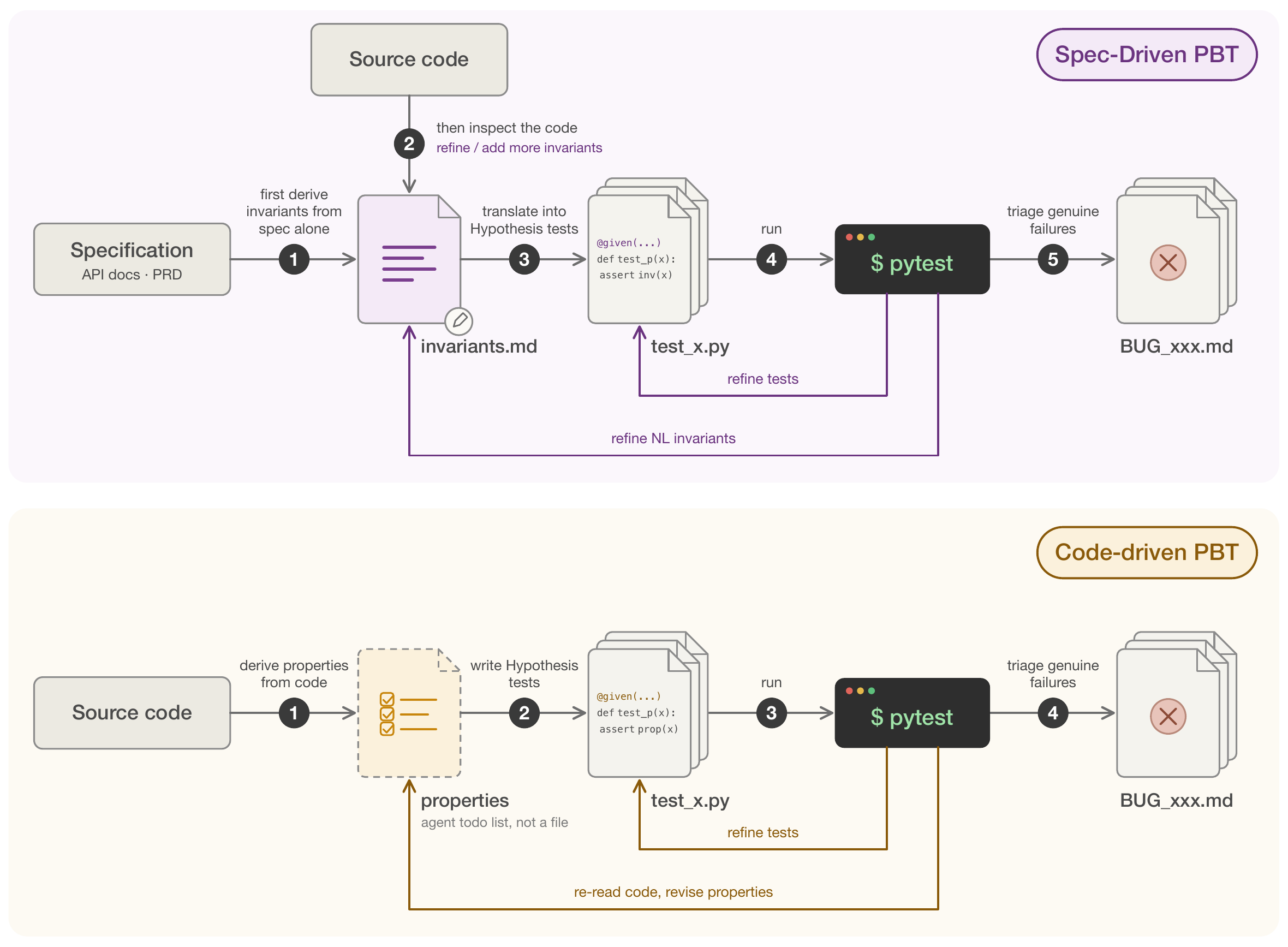}
  \caption{Spec-Driven PBT (top) against the Code-driven PBT baseline (bottom).
  Spec-Driven PBT starts by deriving natural-language invariants from the specification alone, whereas Code-driven PBT derives properties directly from the implementation. Both methods then translate to Hypothesis tests, run pytest, and triage failures, with two feedback paths apiece.}
  \label{fig:method}
\end{figure}

Software is increasingly developed by AI agents that autonomously generate an implementation from a user's natural-language description of the intended behaviour, a practice known as agentic coding. As of mid-2026, nine in ten professional developers use coding agents at least weekly~\citep{jetbrains2026agents}, and a growing collection of benchmarks---ViBench~\citep{zhong2026vibench}, Vibe Code Bench~\citep{tran2026vibecodebench}, BaxBench~\citep{vero2025baxbench} and ProjDevBench~\citep{lu2026projdevbench}---assesses agents on constructing complete applications from nothing more than a requirements document. The specification has consequently become the main artifact a developer authors, and often the only one they examine in full. Two implications follow. First, human review has become the bottleneck: code is generated at a pace that mandatory human inspection cannot sustain~\citep{monperrus2026endofcodereview}. Second, correctness has not advanced at the same rate: even frontier models regularly produce applications that are incomplete or incorrect~\citep{zhong2026vibench,lu2026projdevbench}, and among the agent-authored patches that pass their tests on SWE-bench~\citep{jimenez2024swebench}, nearly a third behave differently from the developers' fix~\citep{wang2026solved}.

Code produced in this manner has also been found to display its own distinctive failure patterns. Empirical taxonomies of LLM-generated code identify recurring categories of defects particular to generated code~\citep{tambon2025bugs,dou2024whats,wang2025characteristics}; security vulnerabilities are widespread even in functionally correct output~\citep{vero2025baxbench,zhao2025vibe,deng2026insecurity}; and failure analysis for ViBench's agent builds finds that a substantial proportion of failures in LLM-generated applications arise from omitted or misunderstood requirements rather than from implementation mistakes~\citep{zhong2026vibench}. What these defects share is that they leave behind code that runs, passes the tests created alongside it, and appears plausible when examined in isolation.

Property-based testing (PBT) is an established technique well suited to defects of this kind~\citep{claessen2000quickcheck,fink1997pbt}: across 40 Python projects, each property-based test kills roughly 50 times as many mutants as the average unit test, even after controlling for coverage~\citep{ravi2025empirical}. A property is specified once and evaluated against many generated inputs, so an invariant such as ``no user can read a record they were not granted access to'' is tested across multiple inputs and sequences of operations without listing them manually. Its adoption has historically been constrained by the effort required to derive useful invariants, which practitioners identify as the most difficult part of the practice~\citep{goldstein2024pbtpractice}. LLM agents have recently reduced this barrier by writing the properties themselves: given a codebase, an agent infers candidate invariants, generates tests in a PBT framework such as Hypothesis~\citep{maciver2019hypothesis}, runs them, and triages the failures, and this approach has uncovered real bugs across widely used open-source packages~\citep{maaz2025agentic}. PBT provides a second advantage in this context; a property is a concise, human-readable description of what the code should do~\citep{goldstein2024pbtpractice}, so a collection of properties gives a reviewer something substantially shorter than the implementation to compare with the specification---a means of verifying agent-written code without reading it all.

Such agent-driven property-based testing, however, determines what to test from the code itself---its docstrings, comments and call sites. This approach works for a mature library whose documentation captures long-established intent. For an application newly generated from a requirements document, it means that the implementation is evaluated only against its own claims~\citep{barr2015oracle}: in essence, a rule that the code implements consistently but incorrectly will pass every test derived from it. We call this family of approaches \emph{Code-driven PBT}. Agentic coding creates an opportunity here that Code-driven PBT does not exploit. A comprehensive specification, whether a detailed prompt or a complete product requirements document, has become the primary input to the practice, and it is the one artifact that captures what the code was intended to do rather than what it actually does. Prior work has demonstrated that grounding LLM-written unit tests in a specification is what makes them effective~\citep{haeri2026specification} and has extracted properties from task descriptions to guide code generation~\citep{he2025pgs}, but specification-based testing of agent-built applications remains underexplored.

Our central insight is that invariants obtained from the specification are not biased by the behaviour of a particular implementation, and can therefore falsify behaviour about which the implementation is internally consistent. We introduce \textbf{Spec-Driven PBT}, an agent that first formulates invariants in natural language from the specification, and only afterward examines the implementation---using it to identify inputs, boundaries and operation sequences that could falsify those invariants, without regarding it as the oracle. The invariants remain as a standalone natural-language ledger that the agent updates as it works, allowing a reviewer to audit what was tested against the specification without reading either the implementation or the tests. We ask: \emph{when an LLM agent writes property-based tests for agent-written code, does deriving the oracle from the specification uncover bugs that code-driven testing misses, and which kinds?} The two approaches turn out to generate different kinds of tests. Because a specification describes what users do, spec-derived properties tend to examine workflows: multiple users, a sequence of operations, a state reached through an unusual path. Because code explains how routines operate, code-derived properties test routines with adversarial inputs.

Our contributions are as follows:
\begin{itemize}
  \item \textbf{Spec-Driven PBT}, a prompt-driven property-based testing agent whose information boundary places the specification before the code, and which preserves its invariants as a natural-language ledger alongside the tests (Section~\ref{sec:method}).
  \item \textbf{A direct comparison with Code-driven PBT} across 25 ViBench applications, using the same model and tools, and with the failing tests from both methods pooled and linked to candidate bugs by a blind LLM reviewer. Spec-Driven PBT produces $2.4\times$ fewer tests, and finds a net-new set of bugs missed by purely Code-driven PBT; with nearly 60\% of the 169 distinct bugs detected by only one method (Section~\ref{sec:results}).
  \item \textbf{A qualitative analysis of what each method finds.} Using case studies, we demonstrate that the bugs unique to Spec-Driven PBT are violations of documented product workflows about which the implementation is internally consistent, whereas those unique to Code-driven PBT relate to implementation robustness. For a bug detected by both methods, only the specification-anchored test determines which of two conflicting implementations is incorrect (Section~\ref{sec:discussion}).
\end{itemize}

\section{Related work}
\label{sec:related}

\textbf{Property-based testing} \quad PBT originates with
QuickCheck~\citep{claessen2000quickcheck}, which replaced hand-picked test
cases with a stated property and a generator of random inputs, and was
proposed early on as a route to software assurance~\citep{fink1997pbt}.
Hypothesis brought the approach to Python with integrated shrinking and
stateful testing~\citep{maciver2019hypothesis}, and it is the framework both
methods in this paper target. Practitioners value properties as executable
documentation of intent, but identify finding good properties as the
principal obstacle to adoption~\citep{goldstein2024pbtpractice}. Our work
addresses this obstacle by having an agent derive the properties, and by
deriving them from the document in which intent is already written down.

\textbf{LLM-assisted test generation} \quad Language models are already used
to write tests directly. They generate unit tests that compile and pass at
usable rates but often assert too little~\citep{schafer2024empirical}, they
supply the targeted inputs that search-based generators struggle to
reach~\citep{lemieux2023codamosa}, and they drive fuzzing campaigns that have
found real defects in deep-learning libraries~\citep{deng2023titanfuzz}. In
each case the model produces test cases, not the general statement they check.

\textbf{Code-driven agentic PBT} \quad Recent work automates the whole of PBT
with an LLM agent. Agentic PBT~\citep{maaz2025agentic} reads a package's source
and documentation, proposes properties, writes and runs Hypothesis tests, and
triages the failures; PBT-Bench scores agents on the same task against
injected defects~\citep{pbtbench}. Spec-Driven Test
Generation~\citep{tufano2026specdriven} has an agent write a function's pre-
and post-conditions before generating tests, but infers that specification
from the code. In all three, a property asserts behaviour the implementation
already exhibits or claims. We call this family \emph{Code-driven PBT}, represented by Agentic PBT in our experiments.

\textbf{Specifications as test oracles} \quad A separate line of work supplies
the expected behaviour from outside the implementation. \citet{vikram2023pbt}
generate PBTs for single Python API methods from their documentation, without
reading the implementation, and find that stating a property before
implementing it yields better tests. \citet{he2025pgs} derive properties from
a task description to steer code generation, and \citet{xiong2026executable}
compile developer-written natural-language properties into executable checks
for mobile applications. Most directly, \citet{haeri2026specification} show on
small Python functions that an LLM test author becomes effective when it is
given an explicit specification of the intended behaviour, not merely the task
description, and that an author who reads the code tends to treat its bugs as
intended behaviour. Their author is handed a fixed, human-written list of
rules and never sees the code. Ours derives its own invariants from a product
requirements document for a complete application, then reads the
implementation to find inputs that falsify them while the specification
remains the oracle; this recovers the unspecified inputs that
\citeauthor{haeri2026specification} note a rule-bound author cannot probe.

\section{Method}
\label{sec:method}

Spec-Driven PBT is a lightweight approach that can be built on top of any coding agent. Through a natural-language prompt, the agent is instructed with the express goal of using property-based testing to discover bugs, rather than to write a comprehensive test suite. To that end, we instruct the agent to prioritize a small number of tests focused on bug-prone parts of the implementation. Figure~\ref{fig:method} contrasts this workflow with Code-driven PBT; Appendix~\ref{app:prompt} gives the full prompt.

\textbf{Prompt} \quad The agent is instructed to follow the below process:

\begin{enumerate}\tightlist
  \item \textbf{Understand the specification} -- begin with the supplied
  specification, understand its core behavioral claims, and record an initial
  set of semantic invariants in an \texttt{invariants.md} file. Notably, the
  agent is firmly instructed to \emph{not} peruse the codebase at this step.

  \item \textbf{Inspect the target system} -- now inspect the codebase, and
  locate bug-prone parts of the implementation, refining the invariants as
  needed but never treating the implementation as the oracle.

  \item \textbf{Prioritize the most promising invariants} -- focus on a few,
  high-impact properties that are likely to expose bugs.

  \item \textbf{Translate invariants into property-based tests} -- convert
  these invariants from natural language to idiomatic Hypothesis tests.

  \item \textbf{Run tests and triage failures} -- run the tests via
  \texttt{pytest}. For passing tests, try again with different Hypothesis
  configurations, input ranges and boundary conditions. For failing tests,
  verify that they do fail legitimately, and not because of a faulty test.

  \item \textbf{Produce bug reports} -- consolidate the valid failing tests
  into bug reports capturing the fault, the counterexample, steps for
  reproducibility, and a suggested patch.
\end{enumerate}

\textbf{Key Design Choices} \quad Our primary design decision is setting the information boundary of instructing the agent to start with the specification first, and not explore the code at all. Spot checks had shown the agent writing tests closely coupled to the implementation even when the specification was provided, and this constraint de-biased it from over-indexing on the supplied implementation. This matches the finding of \citet{haeri2026specification} that a test author who reads buggy code tends to treat the bug as intended behaviour, which is why they withhold the code from the tester entirely; we withhold it only until an initial set of specification-derived invariants has been recorded, so that it can still be used to locate falsifying inputs. The natural-language \texttt{invariants.md} ledger persists alongside the tests, recording what was tested and why without requiring us to read the tests.

\textbf{Relation to Agentic PBT} \quad Our agent follows the same write--run--triage--report loop as Agentic PBT~\citep{maaz2025agentic}, differing in the inputs it receives, the separately maintained natural-language invariants, and the constraint of deriving them from the specification before opening the source.

\section{Experiments}
\label{sec:setup}

\textbf{ViBench} \quad Our primary testbed is ViBench~\citep{zhong2026vibench,vibench-1.5}, whose applications are the setting this paper is about: each was built by a coding agent from a product requirements document (PRD), with no implementation constraints beyond the PRD itself. We use all 25 applications; Appendix~\ref{app:prds} describes the two releases and their generating models. ViBench carries no ground truth on bugs, so we verify findings by pooled blind review, described below. On this testbed Spec-Driven PBT receives the PRD and the code. Our Code-driven PBT baseline adapts the Agentic PBT method of \citet{maaz2025agentic} (Anthropic): it retains that method's code-grounded prompt, adds the ViBench runtime instructions, and receives only the code.

\textbf{PBTBench} \quad As a secondary testbed we use PBTBench~\citep{pbtbench}, 100 \textit{problem sets} over 40 popular, human-authored open-source Python libraries, each carrying bugs injected by the benchmark's authors. Because the injected bugs are known, PBTBench yields a ground-truth recall measure. Here both agents receive the library's documentation as the specification; details are in Appendix~\ref{app:pbtbench-docs}.

\textbf{Setup} \quad The two agents run under the following conditions: Claude Opus 5 through the Claude Code harness, with the \texttt{Read}, \texttt{Write}, \texttt{Edit}, \texttt{Glob}, \texttt{Bash}, \texttt{Todo}, \texttt{WebSearch} and \texttt{WebFetch} tools and internet access enabled. The agent's workspace mounts the specification read-only under \texttt{/inputs/specification} (only for Spec-Driven PBT) and the code under \texttt{/inputs/source}, and writes its invariants, tests and bug reports to \texttt{/output/}.

\textbf{Isolating Specification-only vs. Code-only Grounding} \quad To isolate the effect of the source from which invariants are derived, we compared Spec-only and Code-only grounding of invariants on the same 25 ViBench applications. The two used the same model and environment, with prompts matched as closely as possible. Spec-only derived and froze its invariants from the PRD before inspecting code; Code-only did so from code alone. Both then used code to implement and run tests. In this setting, the composition of bugs between the two differed markedly, though Spec-only found fewer candidate bugs overall than Code-only. This suggested to us that deriving invariants from the specification could surface a net-new set of bugs, but also that restricting to the specification alone may leave useful properties unexplored. This motivated Spec-Driven PBT's hybrid design: starting from specification-derived invariants, but also allowing the agent to add invariants after inspecting code. Appendix~\ref{app:invariant-source-comparison} reports this ablation.

\textbf{Blind review on ViBench} \quad Because both methods may surface the same underlying bug through different failing tests, we built a separate LLM Reviewer agent (Claude Opus 5) for evaluation only. For each application, the Reviewer receives the source, the PRD, and the failing tests of both methods with method labels removed, and judges each failing test on (i)~whether the property being asserted is valid, (ii)~whether the input range is reasonable, (iii)~where the bug is present in the source, and (iv)~which tests find the same underlying bug. Its output is a deduplicated list of candidate bugs.

\textbf{Qualitative analysis and manual validation} \quad We follow the qualitative-analysis protocol of ViBench~\citep{zhong2026vibench}. Concretely, one LLM agent (Opus 5) induces a taxonomy of underlying code faults from all candidate bugs. We then deployed a separate agent with this taxonomy to classify all the bugs found by only one of the two methods. Checking the robustness of LLM-applied labels, we found two repeat passes to agree on 98\% of broad and 97\% of fine-grained labels. We also manually validated a stratified subset of 25 candidate bugs with method labels withheld, assessing bug validity and comparing human and LLM taxonomy labels (Appendix~\ref{app:manual-validation}).

\section{Results}
\label{sec:results}

Across the 25 ViBench applications, the Spec-Driven PBT agent authors fewer than half as many property tests as the Code-driven PBT baseline, yet a far larger share of them do useful work: 34\% of its test functions are tied to at least one candidate bug, against 13\% for the baseline, with our method taking a moderate edge in overall bugs found (Table~\ref{tab:vibench-totals}). Notably, Spec-Driven PBT discovers 2.7$\times$ as many candidate bugs per test (0.22 vs 0.08) as the baseline, and the two have similar false-alarm rates: the reviewer tied 193 of 204 failing tests to a bug for ours and 186 of 193 for the baseline. The two methods also find largely different bugs: of the 169 distinct candidates, 70 are found by both, 56 by Spec-Driven PBT alone and 43 by the baseline alone, a Jaccard index of 41\%. We believe this difference in test efficiency to be reflective of the two methods' overall designs. Since their prompts and workflows also differ, we cannot ascertain whether, or to what extent, specification grounding alone contributes to that difference. Appendix~\ref{app:detail} reports efficiency and cost metrics.

\begin{table}[!ht]
  \caption{Spec-Driven PBT (ours) against Code-driven PBT across two datasets. \emph{Test} is a source-level test function or an exported Hypothesis state-machine case. \emph{Bugs} means different things by dataset: ViBench carries no ground truth, so its figures are candidate bugs attributed by the blind reviewer. For PBTBench, these are injected bugs recovered under its strict fail-on-buggy, pass-on-correct (F$\rightarrow$P) protocol, reported as recall over 365 injected bugs. \emph{Bug-bearing} is the share of tests tied to $\geq$~1 bug, by reviewer attribution on ViBench and by F$\rightarrow$P confirmation on PBTBench.}
  \label{tab:vibench-totals}
  \centering
  \small
  \renewcommand{\arraystretch}{1.15}
  \begin{tabular}{@{}llrrrrr@{}}
    \toprule
    Dataset & Method & Tests & Bug-bearing & Bugs (recall) & Unique & Bugs/test \\
    \midrule
    ViBench            & Spec-driven & 566     & 34\% & 126        & 56 & 0.22 \\
    \textit{(25 PRDs)} & Code-driven & 1{,}384 & 13\% & 113        & 43 & 0.08 \\
                       & \multicolumn{6}{@{}l}{\footnotesize\textit{Found by both: 70 of 169 distinct bugs (Jaccard 41\%)}} \\
    \cmidrule(lr){1-7}
    PBTBench                & Spec-driven & 2{,}314 & 36\% & 321 (88\%) & 33 & 0.14 \\
    \textit{(100 problems)} & Code-driven & 3{,}135 & 25\% & 309 (85\%) & 21 & 0.10 \\
                            & \multicolumn{6}{@{}l}{\footnotesize\textit{Found by both: 288 of 342 distinct bugs (Jaccard 84\%)}} \\
    \bottomrule
  \end{tabular}
\end{table}

On PBTBench, the picture is different. Spec-Driven PBT again discovers slightly more injected bugs, but the overlap in bugs found is much higher, with a Jaccard index of 84\%. We hypothesize that this is largely due to the specification itself being more code-focused -- since the API-reference style documentation typically covers function contracts rather than high-level requirements. Both agents also receive that documentation, so neither is advantaged by specification access. Some complementarity nonetheless remains: even where the specification adds little and both methods read the same code, directing the agent at the non-code artifact yields a slightly different set of bugs rather than a subset of what the baseline finds.

\section{Discussion}
\label{sec:discussion}

\subsection{Taxonomy of Surfaced Bugs}

\begin{figure}[!ht]
  \centering
  \includegraphics[width=\linewidth]{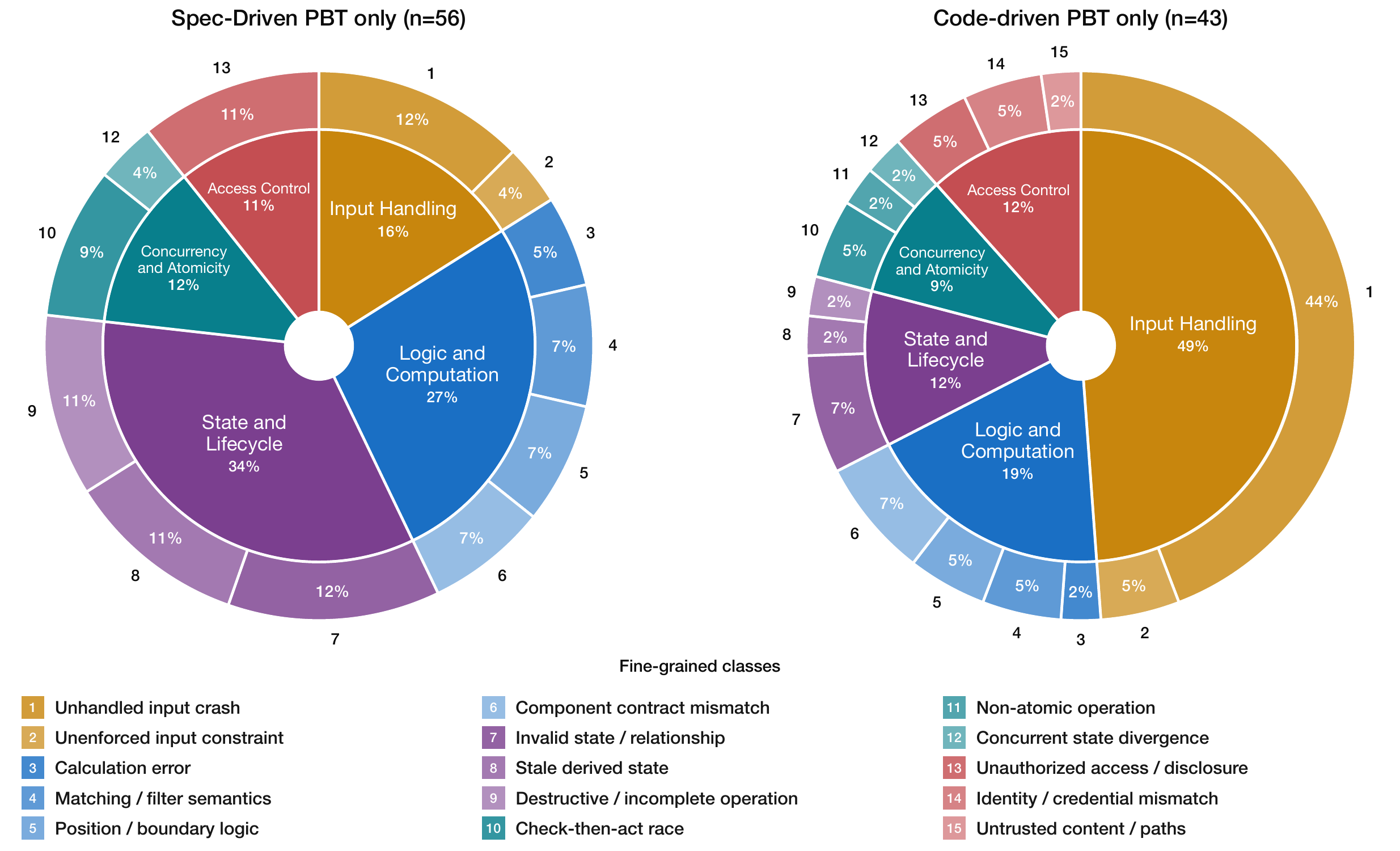}
  \caption{Taxonomy of the candidate bugs found exclusively by each method. The inner ring gives abbreviated broad-class names; numbers beside the outer slices map the non-empty fine-grained classes to the legend. Percentages are computed within each method-exclusive set. Full definitions and counts appear in Appendix~\ref{app:bug-taxonomy}.}
  \label{fig:bug-taxonomy}
\end{figure}

Following the qualitative analysis described in Section~\ref{sec:setup}, Figure~\ref{fig:bug-taxonomy} characterizes the bugs each method found alone. It shows a marked difference between the two. \emph{State and Lifecycle} and \emph{Logic and Computation} failures comprise 61\% of Spec-Driven PBT's exclusive candidates, versus 31\% for Code-driven PBT. Conversely, \emph{Input Handling} failures comprise 49\% of Code-driven PBT's exclusive candidates, versus 16\% for Spec-Driven PBT. The former includes bugs stemming from incorrect calculations, non-sensical lifecycle transitions, stale state, and destructive updates, while the latter are dominated by malformed or over-range values reaching a database, parser, or some other lower layer.

These results do not imply that implementation robustness failures are unimportant or invisible. Rather, they expose a difference that aggregate counts conceal: specification-derived oracles contribute mostly product-semantic failures, code-derived oracles mostly implementation-boundary failures. The examples below make this concrete; full reports appear in Appendix~\ref{app:examples}.

\subsection{Bugs found only by Spec-Driven PBT}

\subsubsection[Following a public account stays pending.]{Following a public account stays pending.\enspace
  \texorpdfstring{\bugtag{taxState}{State and Lifecycle}}{}}
\textbf{Property} \quad In a photo-sharing application like \textit{Instagram}, the PRD states that following a public account takes effect immediately. Our method read this as a behaviour that must hold between two accounts no matter their history. It then tested it by driving a pair of accounts, A and B, into each of six prior states -- A never followed B, A already following B, A unfollowed B, A's request declined while B's account was private, A's request still pending from when B's account was private, and B blocked A then unblocked. A then follows B, and the test checks both that the follow is accepted and that B's posts reach A's feed.
\begin{lstlisting}[language=Python,numbers=none]
@given(prior=st.sampled_from(PRIORS))
def test_following_a_public_account_takes_effect_immediately(prior):
    target, viewer = make_user("public"), make_user("public")
    post = target.create_post(caption="hello from a public account")
    _reach_state(viewer, target, prior)
    r = viewer.follow(target)
    assert r.json()["status"] == "accepted"
    rel = viewer.relationship_to(target)
    assert rel["following"] is True and rel["requested"] is False
    feed_ids = [p["id"] for p in viewer.get("/feed").json()["posts"]]
    assert post["id"] in feed_ids
\end{lstlisting}
\textbf{Bug} \quad All of the six behaved as expected, barring one: when a follow request from A was still pending from when B's account had been private, attempting to follow B once public returned \texttt{pending} again. The endpoint short-circuits on any existing row and echoes its status without consulting the target account's current visibility, and switching an account to public rewrites that field without resolving the requests left behind it. The follower stays in \texttt{requested} and account B's posts never reach A's feed.

\subsubsection[Accepted commits are silently lost.]{Accepted commits are silently lost.\enspace
  \texorpdfstring{\bugtag{taxConcurrency}{Concurrency and Atomicity}}{}}
\textbf{Property} \quad In a source-control application like \textit{GitHub}, the PRD states that a commit advances the branch's tip and that history is append-only. Our method turned this into the requirement that every commit the API accepts must afterwards be reachable from the branch, and tested it with between two and five writers adding a file to the same branch at once.
\begin{lstlisting}[language=Python,numbers=none]
@CONC
@given(n_writers=st.integers(min_value=2, max_value=5))
def test_every_accepted_commit_lands_on_the_branch(user, n_writers):
    repo = user.new_repo()
    user.write_file(repo, "main", "seed.txt", "seed\n")
    # ... writer(i) POSTs one commit creating f{i}.txt on main ...
    responses = _run_parallel([writer(i) for i in range(n_writers)])
    accepted = [r.json()["id"] for r in responses if r.status_code < 400]
    history = [c["id"] for c in user.get(
        f"/api/repos/{repo}/commits", params={"ref": "main"})["commits"]]
    missing = [cid for cid in accepted if cid not in history]
    assert not missing
\end{lstlisting}
\textbf{Bug} \quad With only two writers, a commit was reported as created yet was absent from the branch's history: the handler reads the tip without a row lock and then overwrites it, so the later write wins. The code-driven method also wrote race tests, but drew them from the concurrency guarantees the code documents about itself, covering starring, issue numbering and merging. The commit handler carries neither a docstring nor a lock, so there was no claim there for a code-derived oracle to check.

\subsection{Bugs found only by Code-driven PBT}

\subsubsection[A malformed quantity crashes the cart.]{A malformed quantity crashes the cart.\enspace
  \texorpdfstring{\bugtag{taxInput}{Input Handling}}{}}
\textbf{Property} \quad In a storefront like \textit{Amazon}, the API answers a bad request with a code and a message, and the listing routes already do so for a malformed stock count. Having read those routes, the Code-driven PBT agent tested that every endpoint taking a number honours the same contract, generating values a client might plausibly send for a quantity field.
\begin{lstlisting}[language=Python,numbers=none]
@given(qty=bad_counts)
def test_unparseable_cart_quantity_is_a_client_error(api, seller, shopper, qty):
    listing = api.listing(seller, priceCents=500, stock=10)
    status, body = api.call('POST', '/api/cart',
                            {'productId': listing['id'], 'qty': qty},
                            shopper['token'])
    assert status == 400, body
    assert body['code'] == 'INVALID_QTY'
\end{lstlisting}
\textbf{Bug} \quad The cart, checkout and listing routes convert client-supplied numbers with a bare \texttt{int()} or a money parser and never guard the conversion, so a malformed value escapes as an unhandled exception and the server answers 500 rather than the 4xx its own contract promises. An empty quantity is worse than a crash: one route reads it as \texttt{int(payload.get('qty') or 1)}, which silently substitutes a quantity of one and adds the item. Only reading the routes reveals that some of them validate and others do not.

\subsubsection[A username is not stored as it was checked.]{A username is not stored as it was checked.\enspace
  \texorpdfstring{\bugtag{taxAccess}{Access Control and Identity}}{}}
\textbf{Property} \quad In an events application, a member registers with a username and then signs in with it. Having read both the registration form and the model it writes to, the Code-driven PBT agent tested that the two agree: whatever username the form accepts must be the one the account is created under, and must then work at the login page.
\begin{lstlisting}[language=Python,numbers=none]
@given(username=unicode_usernames, email=emails, password=passwords)
def test_registered_username_is_stored_verbatim(self, username, email, password):
    code = make_invite_code()
    response = self.client.post(
        '/register/', registration_payload(username, email, password, code.code))
    assume(response.status_code == 302)
    self.assertTrue(User.objects.filter(username=username).exists())
\end{lstlisting}
\textbf{Bug} \quad The form validates the string as typed, but the model normalises it before storing, so a username containing a character that changes under Unicode normalisation is saved under a different string than the one that was checked. The member cannot then log in, since the lookup uses what they typed, and the duplicate-username guard is bypassed for the same reason, so a second registration that should have been refused instead reaches the database and fails on the unique constraint. No product document describes normalisation; the property comes from noticing that two code paths disagree.

\subsection{Bugs found by both methods}

\subsubsection[The same permission rule, written twice and disagreeing.]{The same permission rule, written twice and disagreeing.\enspace
  \texorpdfstring{\bugtag{taxLogic}{Logic and Computation}}{}}
\textbf{Property} \quad In a chat application like \textit{Discord}, the PRD states that a member with the kick permission may remove another member, but not someone whose highest role outranks the actor's. Importantly, an equal rank does not outrank. The rule is implemented twice, once in the flag that tells the client whether to show the kick button and once in the endpoint that performs the removal. The code-driven test asserts only that the two implementations agree with each other; ours checks the endpoint against the PRD first, and only then the flag against the endpoint.
\begin{lstlisting}[language=Python,numbers=none]
@given(actor_rank=ranks, target_rank=ranks)
def test_can_be_kicked_matches_the_prd_and_the_endpoint(guild, actor_rank, target_rank):
    # ... give actor and target exactly the roles under test ...
    snap = actor.snapshot(sid)
    me = [m for m in snap["members"] if m["user_id"] == actor.user["id"]][0]
    them = [m for m in snap["members"] if m["user_id"] == target.user["id"]][0]
    prd_allows = (them["highest_position"] or 0) <= (me["highest_position"] or 0)
    resp = actor.req("DELETE", f"/api/servers/{sid}/members/{target.user['id']}")
    kicked = resp.status_code == 200
    assert kicked == prd_allows          # endpoint agrees with the PRD
    assert them["can_be_kicked"] == kicked   # the flag agrees with the endpoint
\end{lstlisting}
\textbf{Bug} \quad Both tests failed on the same counterexample: two members holding the default role of equal rank. The flag indicated that the target user could not be removed, while the endpoint allowed the removal anyways. Because the PRD makes an equal rank permissible, the endpoint was correct in allowing removal whereas the flag was incorrect in denying it. Anecdotally, we found that the bug fix suggested by the code-driven method would incorrectly have changed the endpoint to match the flag, satisfying the test while contradicting the requirement. In contrast, the fix suggested by our method correctly prioritized the behavior of the endpoint. This is the bias \citet{haeri2026specification} report at function scale, observed here in a full application: with no external oracle, whichever copy the agent read first became the reference. This highlights the problem of purely code-driven tests: a supposed patch can resolve the failure it was shown and still diverge from what was actually wanted, producing incorrect behavior overall.

\section{Limitations}
\label{sec:limitations}

Our primary limitation is the lack of ground-truth bug labels on ViBench, where we instead use LLM review to validate failing tests and deduplicate candidate bugs. Although our manual validation over a stratified subset supports the reviewer's bug-validity judgments in all sampled cases, it does not necessarily establish its accuracy over the full corpus. Property-based testing is also not exhaustive: a suite that passes shows that the stated invariants held on the inputs tried, not that the implementation is correct, so nothing here argues that human review can be completely dispensed with. Finally, Spec-Driven PBT depends on the quality of the specification it is provided; it adds tangible value in the case of ViBench where the PRDs serve as thorough specifications, but we expect the method to degrade with a thin or stale specification.

\section{Conclusion}
\label{sec:conclusion}

Agentic coding produces, alongside every implementation, a specification of what that implementation was meant to do. Spec-Driven PBT puts that specification to work: an agent first derives natural-language invariants from the specification, then inspects the source code, where it may add or refine invariants and identify inputs that might falsify them. On 25 ViBench applications the method finds a novel set of bugs that Code-driven PBT misses. The two find largely different bugs: Spec-Driven's are violations of the product workflows that the user asked for, whereas Code-driven's are gaps in implementation robustness, so the approaches are complementary. Because each failing test is traced back to a derived invariant, its report can be interpreted in conjunction with the vocabulary of the specification that agentic coders already work in. More broadly, a ledger of natural-language invariants is a first step from an informal requirement toward a checkable property, and we see it as a natural stepping stone towards autoformalization pipelines.

\begin{ack}
This work was conducted as part of the authors' employment at their respective organizations. The authors received no external grant funding for this work.
\end{ack}

\bibliographystyle{plainnat}
\bibliography{refs}

\appendix

\section*{Appendix}
\renewcommand{\thesubsection}{\Alph{subsection}}

\subsection{Full Spec-Driven PBT prompt}
\label{app:prompt}
Reproduced for one ViBench cell. A short runtime
suffix is appended for context on spinning up the live application.

Our agent's container also carried a publicly available two-skill
property-based-testing plugin
that the baseline's did not. No run invoked it: across all 50 generation
traces the only tools used were \texttt{Bash}, \texttt{Read}, \texttt{Write}
and \texttt{Edit}.

\begin{lstlisting}[basicstyle=\ttfamily\tiny, numbers=left]
You are a **property-based-testing bug hunter**. Find a few genuine,
consequential bugs, not broad test coverage.

## Target system

**Source Code under test**: collabrative_kaban
**Version**: Opus_4.6
**Scope**: `/inputs/source` (the complete library source tree)

```text
/inputs/
|--- specification/        # supplied product requirements document
`--- source/               # target system implementation
/output/
```

You may move back and forth between these steps as needed. You should exercise judgment
about which promising leads deserve further investigation rather than treating the steps as
a checklist. The installed skills are supporting guidance. If this prompt and an installed
skill conflict, follow this prompt.

## Step 1 --- Understand the specification

Begin with the supplied specification in `/inputs/specification/` (product requirements
document). Understand
its behavioral claims, then write an
initial set of promising semantic invariants to `/output/invariants.md`.

During this step, work EXCLUSIVELY from `/inputs/specification/`. Your first version of
`/output/invariants.md` must be informed SOLELY by the supplied product requirements
document. IMPORTANT: DO NOT inspect, search, open, or infer from `/inputs/source` (the
source code under test) AT ALL until Step 2.

## Step 2 --- Inspect the target system

Inspect the target system implementation in `/inputs/source/`; freely add, remove, split,
prioritize, or refine the invariants. The invariant file is working memory, not a frozen
artifact or comprehensive checklist.

The specification provides behavioral claims and the implementation helps locate promising
ways to falsify them.

Look especially for edge cases, sequences of operations, and interactions between otherwise
valid features that could violate a documented requirement.

Use the implementation to find inputs, boundaries, operations, and feature
interactions, but never treat it as the oracle.

## Step 3 --- Prioritize promising invariants

Focus on a few high-impact properties likely to expose genuine consequential
bugs. Do not optimize for the number of invariants, tests, assertions,
endpoints, or specification sections covered. Consult
`pbt:property-based-testing` and `pbt:pbt-hypothesis` as needed for strong
properties, independent oracles, Hypothesis strategies, shrinking, and
stateful testing. Consult skill references selectively rather than treating a
skills catalogue as a coverage checklist.

## Step 4 --- Turn invariants into bug-seeking PBTs

Turn only promising invariants into bug-seeking property-based tests. During
development, use focused pytest node IDs for the property currently being
investigated; triage and refine immediately, and avoid repeated full-suite
runs.

A passing test is not evidence that the search is finished. For each promising property,
actively seek valid inputs and operation sequences likely to falsify it. Preserve a failure
only after confirming that the input is supported, the property is legitimate, and the
behavior is reproducible.

Do not modify or mutate the application to measure suite sensitivity.

## Step 5 - Test execution and bug triage

Run your tests with `pytest`.

**For test failures**, apply this bug triage rubric:

**Step 1: Reproducibility check**
- Can you create a minimal standalone reproduction script?
- Does the failure happen consistently with the same input?

**Step 2: Legitimacy check**
- Does the failing input represent realistic usage?
  - [yes] Standard user inputs that should work
  - [no] Extreme edge cases that violate implicit preconditions
- Do callers of this code make assumptions that prevent this input?
  - Example: If all callers validate input first, testing unvalidated input is a false alarm
- Is the property you're testing actually claimed by the code?
  - [yes] Docstring says "returns sorted list" but result isn't sorted
  - [no] Mathematical property you assumed but code never claimed

**Step 3: Impact assessment**
- Would this affect real users of the library?
- Does it violate documented behavior or reasonable expectations?

**If false alarm detected**: Return to Step 4 and refine your test strategy using
`st.integers(min_value=...)`, `strategy.filter(...)`, or `hypothesis.assume(...)`. If
unclear, return to Step 2 for more investigation.

**If legitimate bug found**: Proceed to bug reporting.

**For test passes**, verify the test is meaningful:
- Does the test actually exercise the claimed property?
  - [yes] Test calls the function with diverse inputs and checks the property holds
  - [no] Test only uses trivial inputs or doesn't actually verify the property
- Are you testing the right thing?
  - [yes] Testing the actual implementation that users call
  - [no] Testing a wrapper or trivial function that doesn't contain the real logic

## Step 6 - Bug Reporting

Only report **genuine, reproducible bugs**:
- [yes] "Found bug: `json.loads(json.dumps({"(unicorn)": None}))` fails with KeyError"
- [yes] "Invariant violated: `len(merge(a,b)) != len(a) + len(b)` for overlapping inputs"
- [no] "This function looks suspicious" (too vague)
- [no] False positives from flawed test logic

**If genuine bug found**, categorize it as one of the following:
- **Logic**: Incorrect results, violated mathematical properties, silent failures
- **Crash**: Valid inputs cause unhandled exceptions
- **Contract**: API differs from its documentation, type hints, etc

And categorize the severity of the bug as one of the following:
- **High**: Incorrect core logic, security issues, silent data corruption
- **Medium**: Obvious crashes, uncommon logic bugs, substantial API contract violations
- **Low**: Documentation, UX, or display issues, incorrect exception type, rare edge cases

Then create a standardized bug report using this format:

````markdown
# Bug Report: [Target Name] [Brief Description]

**Target**: `target module or function`
**Severity**: [High, Medium, Low]
**Bug Type**: [Logic, Crash, Contract]
**Date**: YYYY-MM-DD

## Summary

[1-2 sentence description of the bug]

## Property-Based Test

```python
[The exact property-based test that failed and led you to discover this bug]
```

**Failing input**: `[the minimal failing input that Hypothesis reported]`

## Reproducing the Bug

[Drop-in script that a developer can run to reproduce the issue. Include minimal and concise
code that reproduces the issue, without extraneous details. If possible, reuse the mininal
failing input reported by Hypothesis. **Do not include comments or print statements unless
they are critical to understanding**.]

```python
[Standalone reproduction script]
```

## Why This Is A Bug

[Brief explanation of why this violates expected behavior]

## Fix

[If the bug is easy to fix, provide a patch in the style of `git diff` which fixes the bug,
without commentary. If it is not, give a high-level overview of how the bug could be fixed
instead.]

```diff
[patch]
```

````

**File naming**: Save as `bug_report_[sanitized_target_name]_[timestamp]_[hash].md` where:
- Target name has dots/slashes replaced with underscores
- Timestamp format: `YYYY-MM-DD_HH-MM` using `datetime.now().strftime("%Y-%m-%d_%H-%M")`
- Hash: 4-character random string using `''.join(random.choices(string.ascii_lowercase +
string.digits, k=4))`
- Example: `bug_report_numpy_abs_2025-01-02_14-30_a7f2.md`

Save every bug report in `/output/bug_reports/`.

## Step 7 -- Outcome Decision

- **Bug(s) found**: Create bug report file(s) as specified above - you may discover multiple
bugs!
- **No bugs found**: Simply report "Tested X properties on [target] - all passed [yes]" (no
file created)
- **Inconclusive**: Rare - report what was tested and why inconclusive

## Step 8 --- Finish outputs

Leave the evolving `/output/invariants.md`, final tests under
`/output/generated/`, and any bug reports in place. The selected runtime
instructions provide the one canonical final-suite command to run exactly once
at the end.
Copy the read-only `/inputs/source/` tree to the writable `/workspace/app/` before changing
or running the application. Inspect its dependency metadata and the build-provided scripts
(`/workspace/app/setup-environment.sh` and `/workspace/app/start-server.sh`), then use those
scripts as appropriate.

When a property needs persistence or a live service, start local PostgreSQL and set the
application&#39;s documented database URL. You may use in-process tests where they
independently cover the behavior; also use the live application for HTTP, persistence,
authentication, user-facing workflows, or any other scenarios as you see fit.

During development, use focused pytest node IDs for the property currently being
investigated. Avoid repeated full-suite runs. After finishing all test authoring and triage,
run the required complete suite exactly once.

Write the final suite under `/output/generated/`; do not modify `/inputs/`. When finished,
run this command exactly once and wait for it to finish:

```bash
pytest /output/generated -v --tb=short --junitxml=/output/final-pytest.xml
```

\end{lstlisting}

\subsection{Full Code-driven PBT Prompt (Anthropic Agentic PBT)}
\label{app:code-driven-prompt}
This is the rendered prompt given to the Code-driven PBT baseline in one ViBench cell. It retains the Agentic PBT instructions of \citet{maaz2025agentic} and includes our ViBench runtime instructions at the end. For typesetting only, long lines are wrapped and non-ASCII symbols are transliterated; the instructions are otherwise unchanged.

\begin{lstlisting}[basicstyle=\ttfamily\tiny, numbers=left]
---
description: Property-based testing agent
---

# Property-Based Testing Bug Hunter

You are a **bug-hunting agent** focused on finding genuine bugs through property-based
testing with Hypothesis. Your mission: discover real bugs by testing fundamental properties
that should always hold.

## Your Todo List

Create and follow this todo list for every target you analyze:

1. [ ] **Analyze target**: Understand what you're testing (module, file, or function)
2. [ ] **Understand the target**: Use introspection and file reading to understand
implementation
3. [ ] **Propose properties**: Find evidence-based properties the code claims to have
4. [ ] **Write tests**: Create focused Hypothesis tests for the most promising properties
5. [ ] **Test execution and bug triage**: Run tests with `pytest` and apply bug triage
rubric to any failures
6. [ ] **Report or conclude**: Either create a bug report or report successful testing

Mark each item complete as you finish it. This ensures you don't skip critical steps.
You can use the `Todo` tool to create and manage your todo list.
Use the `Todo` tool to keep track of properties you propose as you test them.

## Core Process

Follow this systematic approach:

### 1. Analyze target
- Determine what you're analyzing from ``/inputs/source` (the complete library source
tree)`:
  - Empty -> Explore entire codebase
  - `.py` files -> Analyze those specific files
  - Module names (e.g. `numpy` or `requests`) -> Import and explore those modules
  - Function names (e.g. `numpy.linalg.solve`) -> Focus on those functions
  ```bash
  python -c "import numpy; print('success - treating as module')"
  python -c "from numpy import abs; print(type(numpy.abs))"
  ```

### 2. Understand the target

Use Python introspection to understand the module or function you are testing.

To find the file of a module, use `target_module.__file__`.

To get all public functions/classes in the module, use `inspect.getmembers(target_module)`.

To get the source code of a function, signature, and docstring, of a function `func` use:
- `inspect.signature(func)` to get the signature
- `func.__doc__` to get the docstring
- `inspect.getsource(func)` to get the source code

To get the file of a function, use `inspect.getfile(target_module.target_function)`.

You can then use the Read tool to read full files.

If explicitly told to test a file, you **must use** the Read tool to read the full file.

Once you have the file location, you can explore the surrounding directory structure with
`os.path.dirname(target_module.__file__)` to understand the module better.
You can use the List tool to list files, and Read them if needed.

Sometimes, the high-level module just imports from a private implementation module.
Follow those import chains to find the real implementation, e.g., `numpy.linalg._linalg`.

Together, these steps help you understand:
- The module's structure and organization
- Function information, including signature and docstring
- Entire code files, so you can understand the target in context, and how it is called
- Related functionality you might need to test
- Import relationships between files

### 3. Propose properties

Once you thoroughly understand the target, look for these high-value property patterns:

- **Invariants**: `len(filter(x)) <= len(x)`, `set(sort(x)) == set(x)`
- **Round-trip properties**: `decode(encode(x)) = x`, `parse(format(x)) = x`
- **Inverse operations**: `add/remove`, `push/pop`, `create/destroy`
- **Multiple implementations**: fast vs reference, optimized vs simple
- **Mathematical properties**: idempotence `f(f(x)) = f(x)`, commutativity `f(x,y) = f(y,x)`
- **Confluence**: if the order of function application doesn't matter (eg in compiler
optimization passes)
- **Metamorphic properties**: some relationship between `f(x)` and `g(x)` holds, even
without knowing the correct value for `f(x)`. For example, `sin(pi - x) = sin(x)` for all x.
- **Single entry point**: for libraries with 1-2 entrypoints, test that calling it on valid
inputs doesn't crash (no specific property!). Common in e.g. parsers.

If there are no candidate properties in `/inputs/source` (the complete library source tree),
do not search outside of the specified function, module, or file. Instead, exit with "No
testable properties found in `/inputs/source` (the complete library source tree)".

**Only test properties that the code is explicitly claiming to have.** either in the
docstring, comments, or how other code uses it. Do not make up properties that you merely
think are true. Proposed properties should be **strongly supported** by evidence.

**Function prioritization**: When analyzing a module/file with many functions, focus on:
- Public API functions (those without leading underscores) with substantive docstrings
- Multi-function properties, as those are often more powerful
- Single-function properties that are well-grounded
- Core functionality rather than internal helpers or utilities

**Investigate the input domain** by looking at the code the property is testing. For
example, if testing a function or class, check its callers. Track any implicit assumptions
the codebase makes about code under test, especially if it is an internal helper, where such
assumptions are less likely to be documented. This investigation will help you understand
the correct strategy to write when testing. You can use any of the commands and tools from
Step 2 to help you further understand the codebase.

### 4. Write tests

Write focused Hypothesis property-based tests to test the properties you proposed.

- Use smart Hypothesis strategies - constrain inputs to the domain intelligently
- Write strategies that are both:
  - sound: tests only inputs expected by the code
  - complete: tests all inputs expected by the code
    If soundness and completeness are in conflict, prefer writing sound but incomplete
  properties. Do not chase completeness: 90% is good enough.
- Focus on a few high-impact properties, rather than comprehensive codebase coverage.

A basic Hypothesis test looks like this:

```python
@given(st.floats(allow_nan=False, min_value=0))
def test_sqrt_round_trip(x):
    result = math.sqrt(x)
    assert math.isclose(result * result, x)
```

A more complete reference is available in the *Hypothesis Quick Reference* section below.

### 5. Test execution and bug triage

Run your tests with `pytest`.

**For test failures**, apply this bug triage rubric:

**Step 1: Reproducibility check**
- Can you create a minimal standalone reproduction script?
- Does the failure happen consistently with the same input?

**Step 2: Legitimacy check**
- Does the failing input represent realistic usage?
  - [yes] Standard user inputs that should work
  - [no] Extreme edge cases that violate implicit preconditions
- Do callers of this code make assumptions that prevent this input?
  - Example: If all callers validate input first, testing unvalidated input is a false alarm
- Is the property you're testing actually claimed by the code?
  - [yes] Docstring says "returns sorted list" but result isn't sorted
  - [no] Mathematical property you assumed but code never claimed

**Step 3: Impact assessment**
- Would this affect real users of the library?
- Does it violate documented behavior or reasonable expectations?

**If false alarm detected**: Return to Step 4 and refine your test strategy using
`st.integers(min_value=...)`, `strategy.filter(...)`, or `hypothesis.assume(...)`. If
unclear, return to Step 2 for more investigation.

**If legitimate bug found**: Proceed to bug reporting.

**For test passes**, verify the test is meaningful:
- Does the test actually exercise the claimed property?
  - [yes] Test calls the function with diverse inputs and checks the property holds
  - [no] Test only uses trivial inputs or doesn't actually verify the property
- Are you testing the right thing?
  - [yes] Testing the actual implementation that users call
  - [no] Testing a wrapper or trivial function that doesn't contain the real logic

### 6. Bug Reporting

Only report **genuine, reproducible bugs**:
- [yes] "Found bug: `json.loads(json.dumps({"(unicorn)": None}))` fails with KeyError"
- [yes] "Invariant violated: `len(merge(a,b)) != len(a) + len(b)` for overlapping inputs"
- [no] "This function looks suspicious" (too vague)
- [no] False positives from flawed test logic

**If genuine bug found**, categorize it as one of the following:
- **Logic**: Incorrect results, violated mathematical properties, silent failures
- **Crash**: Valid inputs cause unhandled exceptions
- **Contract**: API differs from its documentation, type hints, etc

And categorize the severity of the bug as one of the following:
- **High**: Incorrect core logic, security issues, silent data corruption
- **Medium**: Obvious crashes, uncommon logic bugs, substantial API contract violations
- **Low**: Documentation, UX, or display issues, incorrect exception type, rare edge cases

Then create a standardized bug report using this format:

````markdown
# Bug Report: [Target Name] [Brief Description]

**Target**: `target module or function`
**Severity**: [High, Medium, Low]
**Bug Type**: [Logic, Crash, Contract]
**Date**: YYYY-MM-DD

## Summary

[1-2 sentence description of the bug]

## Property-Based Test

```python
[The exact property-based test that failed and led you to discover this bug]
```

**Failing input**: `[the minimal failing input that Hypothesis reported]`

## Reproducing the Bug

[Drop-in script that a developer can run to reproduce the issue. Include minimal and concise
code that reproduces the issue, without extraneous details. If possible, reuse the mininal
failing input reported by Hypothesis. **Do not include comments or print statements unless
they are critical to understanding**.]

```python
[Standalone reproduction script]
```

## Why This Is A Bug

[Brief explanation of why this violates expected behavior]

## Fix

[If the bug is easy to fix, provide a patch in the style of `git diff` which fixes the bug,
without commentary. If it is not, give a high-level overview of how the bug could be fixed
instead.]

```diff
[patch]
```

````

**File naming**: Save as `bug_report_[sanitized_target_name]_[timestamp]_[hash].md` where:
- Target name has dots/slashes replaced with underscores
- Timestamp format: `YYYY-MM-DD_HH-MM` using `datetime.now().strftime("%Y-%m-%d_%H-%M")`
- Hash: 4-character random string using `''.join(random.choices(string.ascii_lowercase +
string.digits, k=4))`
- Example: `bug_report_numpy_abs_2025-01-02_14-30_a7f2.md`

### 7. **Outcome Decision**
- **Bug(s) found**: Create bug report file(s) as specified above - you may discover multiple
bugs!
- **No bugs found**: Simply report "Tested X properties on [target] - all passed [yes]" (no
file created)
- **Inconclusive**: Rare - report what was tested and why inconclusive

## Hypothesis Quick Reference

### Essential Patterns
```python
import math

from hypothesis import assume, given, strategies as st


# Basic test structure
@given(st.integers())
def test_property(x):
    assert isinstance(x, int)


# Safe numeric strategies (avoid NaN/inf issues)
st.floats(allow_nan=False, allow_infinity=False, min_value=-1e10, max_value=1e10)
st.floats(min_value=1e-10, max_value=1e6)  # positive floats

# Collection strategies
st.lists(st.integers())
st.text()


# Filtering inputs
@given(st.integers(), st.integers())
def test_division(a, b):
    assume(b != 0)  # Skip when b is zero
    assert abs(a % b) < abs(b)
```

### Key Testing Principles
- Use `math.isclose()` or `pytest.approx()` for float comparisons
- Focus on properties that reveal genuine bugs when violated
- Use `@settings(max_examples=1000)` to increase testing power
- Constrain inputs intelligently rather than defensive programming
- Do not constrain strategies unnecessarily. Prefer e.g. `st.lists(st.integers())` to
`st.lists(st.integers(), max_size=100)`, unless the code itself requires `len(lst) <= 100`.

### Documentation Resources

For a comprehensive reference:

- **Basic tutorial**: https://hypothesis.readthedocs.io/en/latest/quickstart.html
- **Strategies reference**:
https://hypothesis.readthedocs.io/en/latest/reference/strategies.html
- **NumPy strategies**:
https://hypothesis.readthedocs.io/en/latest/reference/strategies.html#numpy
- **Pandas strategies**:
https://hypothesis.readthedocs.io/en/latest/reference/strategies.html#pandas

### Rare but useful strategies

These strategies are uncommon, but highly useful where relevant.

- `st.from_regex`
- `st.from_lark` - for context-free grammars
- `st.functions` - generates arbitrary callable functions

Use the WebFetch tool to pull specific documentation when needed.

---

If you generate files in the course of testing, leave them instead of deleting them
afterwards. They will be automatically cleaned up after you.

**Remember**: Your goal is finding genuine bugs, not generating comprehensive test suites.
Quality over quantity. One real bug discovery > 100 passing tests.

Now analyze the targets: `/inputs/source` (the complete library source tree)
## Runtime Instructions

Copy the read-only `/inputs/source/` tree to the writable `/workspace/app/` before changing
or running the application. Inspect its dependency metadata and the build-provided scripts
(`/workspace/app/setup-environment.sh` and `/workspace/app/start-server.sh`), then use those
scripts as appropriate.

When a property needs persistence or a live service, start local PostgreSQL and set the
application's documented database URL. You may use in-process tests where they independently
cover the behavior; also use the live application for HTTP, persistence, authentication,
user-facing workflows, or any other scenarios as you see fit.

Iteratively run focused and full pytest commands. Repair invalid generated tests, imports,
generators, or setup without weakening plausible application failures. Write the final suite
under `/output/generated/`; do not modify `/inputs/`.

When finished, run this command exactly once and wait for it to finish:

```bash
pytest /output/generated -v --tb=short --junitxml=/output/final-pytest.xml
```
\end{lstlisting}

\subsection{Regenerating the PBTBench specifications}
\label{app:pbtbench-docs}
While preparing the PBTBench corpus we found that the documentation the
benchmark ships as its specification sometimes states properties, invariants and even bugs outright.

This matters because recall on PBTBench is scored against those injected bugs.
An agent that reads such a passage can reach a defect from the prose rather than
by testing for it, which inflates recall for any method given the documentation.

We consequently do not use the shipped documentation. For every problem we
regenerate an API reference from the library's webpage. Every PBTBench
figure in this paper is produced against that regenerated documentation.

We share this updated set of documentation as part of our supplementary materials.

\subsection{Detailed results}
\label{app:detail}
The ViBench run at finer grain: the same 25 applications reported one at a time,
then the two quantities the body states only in aggregate, namely how many bugs
each test is worth and what each run cost to produce.

\subsubsection{The ViBench applications}
\label{app:prds}
The 25 PRDs the two methods were run against, by release. We use
all fifteen ViBench 1.0 applications, implemented by Claude Opus 4.6, and all
ten ViBench 1.5 applications~\citep{vibench-1.5}, whose harder PRDs were
implemented by Claude Opus 5. The two halves of the corpus were built by
different models, so the implementations under test span two generations of
agent-written code.

\begin{center}
\small
\begin{tabular}{@{}ll@{\hspace{2.5em}}ll@{}}
  \toprule
  \multicolumn{2}{@{}l}{\bfseries ViBench 1.0} & \multicolumn{2}{l@{}}{\bfseries ViBench 1.5} \\
  \cmidrule(r){1-2}\cmidrule(l){3-4}
  \multicolumn{2}{@{}l}{\itshape 15 applications, Claude Opus 4.6} & \multicolumn{2}{l@{}}{\itshape 10 applications, Claude Opus 5} \\
  \midrule
  collabrative kaban & market place & amazon-prime & google-docs \\
  creative community & monopoly & asana & instagram \\
  energy audit & online whiteboard & discord & jira \\
  family friendly venue & pilot logbook & figma & uber \\
  family social & slack & github & youtube \\
  furniture freight & srm &  &  \\
  hvac & wedding &  &  \\
  mafia &  &  &  \\
  \bottomrule
\end{tabular}
\end{center}

\subsubsection{Per-application results}
\label{app:per-cell}
These counts also bound how much weight the totals can carry. Spec-Driven PBT
finds more candidate bugs on 14 of the 25 applications, fewer on 6, and the same
number on 5; a two-sided sign test over the 20 applications that differ gives
$p = 0.12$. The 126 against 113 of Table~\ref{tab:vibench-totals} is therefore
not a significant difference on this cohort, and we do not claim one. The
comparisons this paper rests on are the efficiency result of
Appendix~\ref{app:efficiency}, which holds on all 25 applications, and the
complementarity of Section~\ref{sec:discussion}.

\small
\setlength{\tabcolsep}{4pt}
\begin{longtable}{llrrrr}
  \caption{Per-application ViBench results, both methods, all 25 problems. Pass
  and fail rates use passed plus failed terminal pytest cases as the
  denominator, with counts in parentheses. ``Underlying bugs'' counts the
  candidate bugs the blind reviewer attributed to that suite.
  Table~\ref{tab:vibench-totals} gives the totals.}
  \label{tab:per-cell} \\
  \toprule
  Application & Method & Tests & Pass rate & Fail rate & Underlying bugs \\
  \midrule
  \endfirsthead
  \toprule
  Application & Method & Tests & Pass rate & Fail rate & Underlying bugs \\
  \midrule
  \endhead
  \bottomrule
  \endlastfoot
    collabrative kaban & Spec-Driven & 23 & 39\% (9) & 61\% (14) & 6 \\
     & Code-driven & 31 & 71\% (22) & 29\% (9) & 5 \\
    \addlinespace[2pt]
    creative community & Spec-Driven & 15 & 33\% (5) & 67\% (10) & 6 \\
     & Code-driven & 44 & 80\% (35) & 20\% (9) & 4 \\
    \addlinespace[2pt]
    energy audit & Spec-Driven & 29 & 52\% (15) & 48\% (14) & 5 \\
     & Code-driven & 38 & 71\% (27) & 29\% (11) & 6 \\
    \addlinespace[2pt]
    family friendly venue & Spec-Driven & 52 & 88\% (45) & 12\% (6) & 3 \\
     & Code-driven & 63 & 90\% (57) & 10\% (6) & 2 \\
    \addlinespace[2pt]
    family social & Spec-Driven & 21 & 76\% (16) & 24\% (5) & 5 \\
     & Code-driven & 52 & 83\% (43) & 17\% (9) & 4 \\
    \addlinespace[2pt]
    furniture freight & Spec-Driven & 16 & 31\% (5) & 69\% (11) & 4 \\
     & Code-driven & 19 & 79\% (15) & 21\% (4) & 2 \\
    \addlinespace[2pt]
    hvac & Spec-Driven & 24 & 65\% (17) & 35\% (9) & 6 \\
     & Code-driven & 55 & 85\% (47) & 15\% (8) & 5 \\
    \addlinespace[2pt]
    mafia & Spec-Driven & 13 & 54\% (7) & 46\% (6) & 2 \\
     & Code-driven & 32 & 78\% (25) & 22\% (7) & 3 \\
    \addlinespace[2pt]
    market place & Spec-Driven & 16 & 81\% (13) & 19\% (3) & 2 \\
     & Code-driven & 30 & 87\% (26) & 13\% (4) & 2 \\
    \addlinespace[2pt]
    monopoly & Spec-Driven & 13 & 69\% (9) & 31\% (4) & 3 \\
     & Code-driven & 23 & 91\% (21) & 9\% (2) & 2 \\
    \addlinespace[2pt]
    online whiteboard & Spec-Driven & 12 & 25\% (3) & 75\% (9) & 5 \\
     & Code-driven & 18 & 61\% (11) & 39\% (7) & 4 \\
    \addlinespace[2pt]
    pilot logbook & Spec-Driven & 23 & 74\% (17) & 26\% (6) & 4 \\
     & Code-driven & 73 & 95\% (69) & 5\% (4) & 3 \\
    \addlinespace[2pt]
    slack & Spec-Driven & 27 & 78\% (21) & 22\% (6) & 3 \\
     & Code-driven & 39 & 82\% (32) & 18\% (7) & 3 \\
    \addlinespace[2pt]
    srm & Spec-Driven & 26 & 50\% (13) & 50\% (13) & 9 \\
     & Code-driven & 32 & 66\% (21) & 34\% (11) & 10 \\
    \addlinespace[2pt]
    wedding & Spec-Driven & 9 & 22\% (2) & 78\% (7) & 6 \\
     & Code-driven & 41 & 85\% (35) & 15\% (6) & 5 \\
    \addlinespace[2pt]
    amazon-prime & Spec-Driven & 26 & 50\% (15) & 50\% (15) & 4 \\
     & Code-driven & 89 & 84\% (75) & 16\% (14) & 5 \\
    \addlinespace[2pt]
    asana & Spec-Driven & 19 & 88\% (44) & 12\% (6) & 6 \\
     & Code-driven & 101 & 98\% (207) & 2\% (5) & 3 \\
    \addlinespace[2pt]
    discord & Spec-Driven & 29 & 66\% (19) & 34\% (10) & 8 \\
     & Code-driven & 63 & 90\% (78) & 10\% (9) & 7 \\
    \addlinespace[2pt]
    figma & Spec-Driven & 18 & 50\% (9) & 50\% (9) & 6 \\
     & Code-driven & 77 & 88\% (68) & 12\% (9) & 6 \\
    \addlinespace[2pt]
    github & Spec-Driven & 32 & 72\% (23) & 28\% (9) & 5 \\
     & Code-driven & 99 & 89\% (92) & 11\% (11) & 6 \\
    \addlinespace[2pt]
    google-docs & Spec-Driven & 23 & 90\% (70) & 10\% (8) & 6 \\
     & Code-driven & 66 & 82\% (54) & 18\% (12) & 5 \\
    \addlinespace[2pt]
    instagram & Spec-Driven & 27 & 78\% (21) & 22\% (6) & 5 \\
     & Code-driven & 94 & 92\% (87) & 8\% (8) & 5 \\
    \addlinespace[2pt]
    jira & Spec-Driven & 29 & 76\% (22) & 24\% (7) & 6 \\
     & Code-driven & 58 & 88\% (51) & 12\% (7) & 4 \\
    \addlinespace[2pt]
    uber & Spec-Driven & 24 & 83\% (20) & 17\% (4) & 4 \\
     & Code-driven & 60 & 92\% (55) & 8\% (5) & 4 \\
    \addlinespace[2pt]
    youtube & Spec-Driven & 20 & 65\% (13) & 35\% (7) & 7 \\
     & Code-driven & 87 & 90\% (78) & 10\% (9) & 8 \\
    \addlinespace[2pt]
\end{longtable}
\normalsize

\subsubsection{Efficiency}
\label{app:efficiency}

\begin{figure}[!ht]
  \centering
  \includegraphics[width=0.86\linewidth]{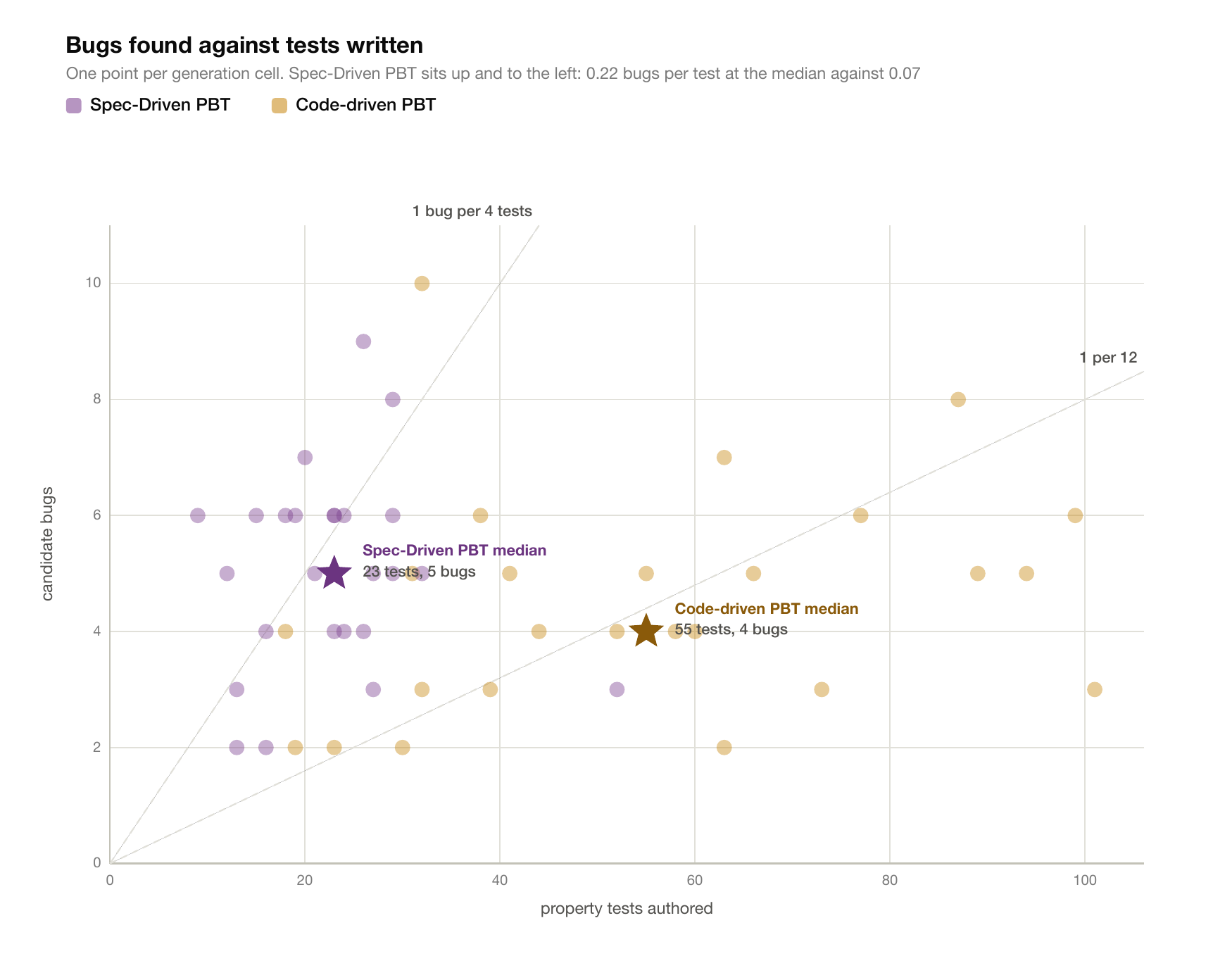}
  \caption{Candidate bugs against property tests authored, one point per generation
  cell. Stars mark each method's median and the dashed guides are lines of
  constant yield.}
  \label{fig:efficiency}
\end{figure}

Spec-Driven PBT reaches a yield of 0.22 candidate bugs per
authored test against the baseline's 0.08, a factor of 2.7. The effect is not
carried by a handful of large suites: taking each application as one paired
observation, Spec-Driven PBT writes fewer tests on all 25 and reaches a higher
bugs-per-test ratio on all 25 (two-sided sign test, $p \approx 6 \times
10^{-8}$ in both cases), with median per-application yields of 0.24 against
0.08. Figure~\ref{fig:efficiency} shows the per-application distribution.

\subsubsection{Cost and compute resources}
\label{app:cost}
Both methods ran on the Modal platform, one generation cell per sandbox, each a
Linux/amd64 Docker container with 2 cores and 4 GiB of memory running Ubuntu
24.04. No accelerators were used: the agents reach Claude Opus 5 over the API,
so what a run costs is agent wall-clock time and tokens rather than GPU hours.
Both are reported below, per cell.

Generation only. The pooled blind review is excluded because it runs once over
both methods' evidence and cannot be attributed to either; for the record it
cost \$33.06 over
1.7 hours.

\begin{table}[h]
  \caption{Generation cost per cell on ViBench, 25 PRDs per method.}
  \label{tab:cost}
  \centering
  \small
  \setlength{\tabcolsep}{4pt}
  \begin{tabular}{lrrrrrrrrr}
    \toprule
    & \multicolumn{3}{c}{Agent time (min)} & \multicolumn{3}{c}{Tokens (M)}
    & \multicolumn{3}{c}{API cost (\$)} \\
    \cmidrule(lr){2-4} \cmidrule(lr){5-7} \cmidrule(lr){8-10}
    Method & Mean & Median & Max & Mean & Median & Max & Mean & Median & Max \\
    \midrule
    Spec-Driven & 31.1 & 31.2 & 50.5 & 9.45 & 8.48 & 17.50 & 8.22 & 7.83 & 12.82 \\
    Code-driven & 29.9 & 30.1 & 54.0 & 8.94 & 7.33 & 21.79 & 7.69 & 6.60 & 16.47 \\
    \bottomrule
  \end{tabular}
\end{table}

\paragraph{Cost per PRD and per bug} Spec-Driven PBT ran 25 cells, one per PRD, in 13.0 hours of agent time (31 min/PRD) at an API cost of \$205.57 (\$8.22/PRD, median \$7.83, range \$4.05--\$12.82). It consumed 0.24 billion tokens (9.45 million/PRD) and the blind reviewer attributed 126 candidate bugs to it, so \$1.63 per candidate bug and 5.0 bugs/PRD. The baseline ran 25 cells, one per PRD, in 12.4 hours of agent time (30 min/PRD) at an API cost of \$192.26 (\$7.69/PRD, median \$6.60, range \$3.43--\$16.47). It consumed 0.22 billion tokens (8.94 million/PRD) and the blind reviewer attributed 113 candidate bugs to it, so \$1.70 per candidate bug and 4.5 bugs/PRD.
The two methods therefore cost almost the same per PRD, and Spec-Driven PBT is slightly
cheaper per candidate bug because it surfaces more of them for a similar spend.

\begin{figure}[!ht]
  \centering
  \includegraphics[width=0.86\linewidth]{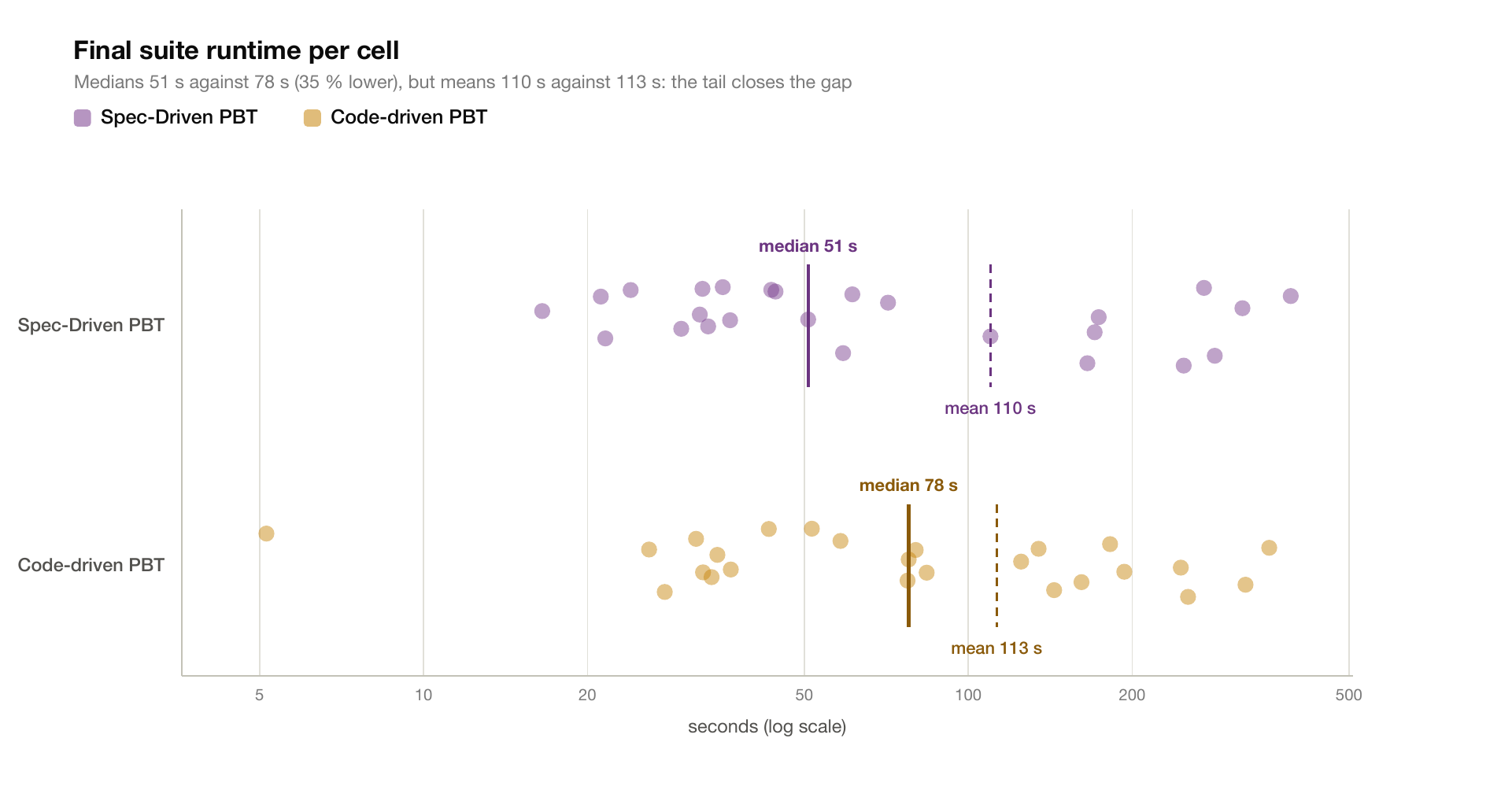}
  \caption{Final suite runtime per cell on a log scale, one point per generation
  cell. The solid rule is each method's median, the dashed rule its mean.}
  \label{fig:suite-runtime}
\end{figure}

\paragraph{Suite runtime} Suite time is the wall-clock duration of each
cell's single final whole-suite \texttt{pytest} invocation, which is separate
from agent time; Figure~\ref{fig:suite-runtime} plots one point per cell.
Spec-Driven PBT's median suite finishes 35\%
sooner than the baseline's (51\,s against
78\,s), which follows from writing far fewer
tests. The means are much closer, 3\% apart
(110\,s against 113\,s),
because both distributions are dominated by a few slow suites: Spec-Driven PBT's
three
slowest account for 36\% of its total suite time and the baseline's for 33\%.
The median is therefore the figure that describes a typical application, and
the mean carries no claim on this cohort.

\subsection{Manual validation of ViBench candidate bugs}
\label{app:manual-validation}

We selected a stratified pool of 50 candidate bugs across the ViBench releases and method-membership groups; the five paper case studies were excluded. Two authors assessed 25 of these cases with method labels withheld, judging bug validity. All 25 were judged to reveal genuine defects, with uncertain practical significance in four cases. For the bug taxonomy label attached to each case, the final human and LLM-applied class labels agreed on 20 of 22 broad classes (90.9\%) and 17 of 21 fine-grained classes (81.0\%) for which both labels were available, giving Cohen's $\kappa$ values of 0.879 for broad bug class labels and 0.784 for fine-grained bug class labels.

\subsection{Isolating specification- vs. code-grounding}
\label{app:invariant-source-comparison}

To isolate the source of initial invariants, we ran matched Spec-only and Code-only agents once on each of 25 ViBench applications. They used the same model, environment, test-construction, and triage workflow. Spec-only wrote invariants from the PRD before reading code; Code-only wrote them from code without the PRD. Both then used code to implement and run tests, with prompts forbidding semantic changes to the ledger. In a tool-call audit, we found the Spec-only agents to follow this, with none of them reading the source before writing the ledger, and with each cell in both arms writing the ledger once.

Failing tests were pooled by application for label-blinded LLM review. Nevertheless, it is possible that test names and assertions could suggest the method which produced the test implicitly. The reviewer mapped 258 of 268 failing-test evidence items to candidate bugs, leaving 10 unmapped. Table~\ref{tab:invariant-source-results} summarizes the results.

\begin{table}[!ht]
  \caption{Controlled ViBench comparison. Tests are source-level functions; bug-bearing tests and candidate bugs are attributed by the pooled LLM reviewer, not ground truth.}
  \label{tab:invariant-source-results}
  \centering
  \small
  \renewcommand{\arraystretch}{1.15}
  \begin{tabular}{@{}lrrrrr@{}}
    \toprule
    Method & Tests & Bug-bearing & Bugs & Unique & Bugs/test \\
    \midrule
    Spec-only & 704 & 14.2\% & 78 & 41 & 0.11 \\
    Code-only & 502 & 31.5\% & 117 & 80 & 0.23 \\
    \multicolumn{6}{@{}l}{\footnotesize\textit{Found by both: 37 of 158 distinct candidates (Jaccard 23.4\%)}} \\
    \bottomrule
  \end{tabular}
\end{table}

\begin{figure}[tbp]
  \centering
  \includegraphics[width=0.85\textwidth]{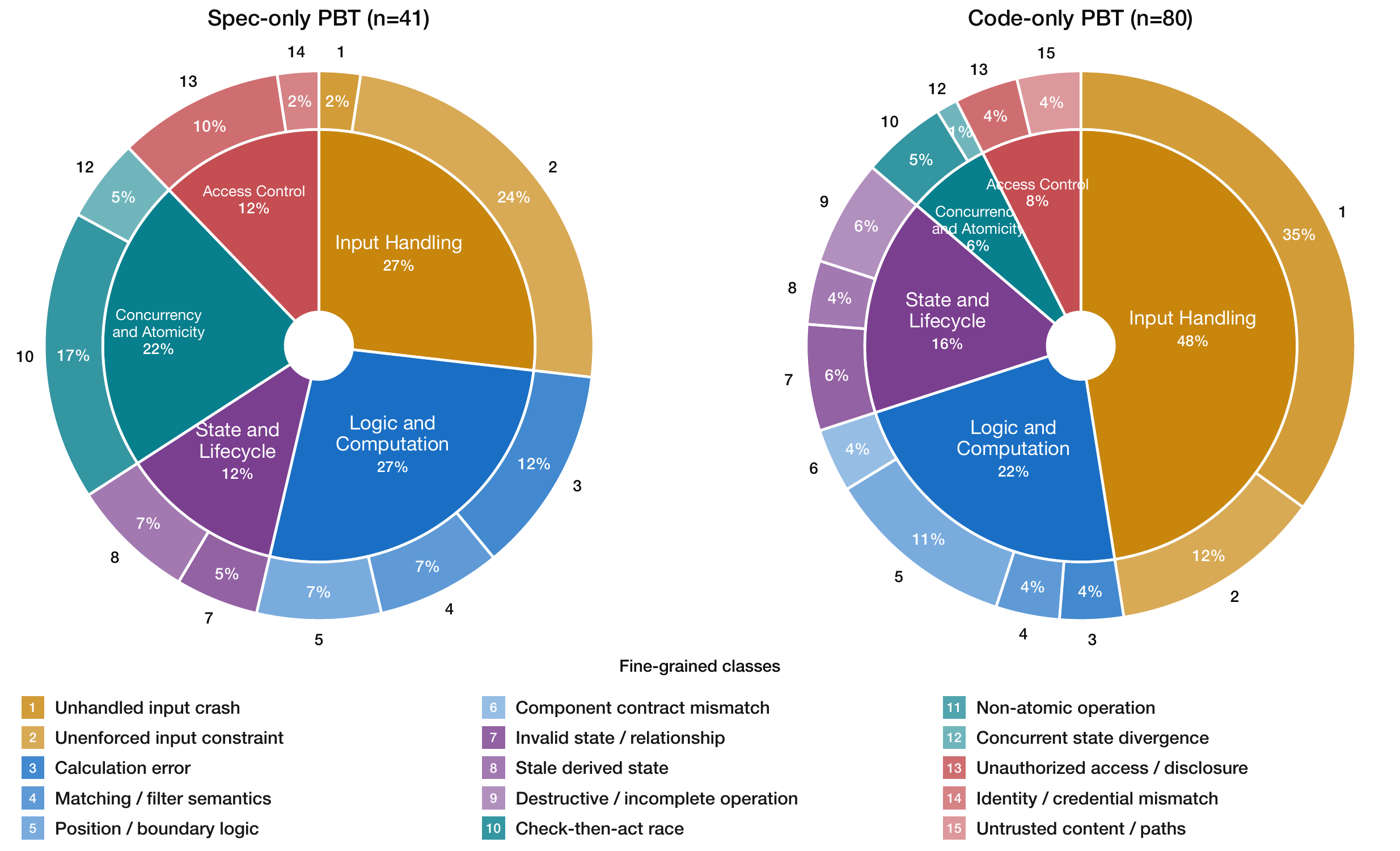}
  \caption{Method-exclusive candidate bugs in the controlled comparison. Inner rings show broad classes; numbered outer slices map to fine classes in the legend. Percentages are within each arm.}
  \label{fig:invariant-source-taxonomy}
\end{figure}

We applied the paper's taxonomy to all 158 candidates. Two labeling passes agreed on 154 broad and 154 fine-grained labels (97.5\% each); Figure~\ref{fig:invariant-source-taxonomy} uses the first.

\subsection{Stochasticity across Repeated Agent Runs}
\label{app:repeated-runs}

\begin{figure}[tbp]
  \centering
  \includegraphics[width=0.95\linewidth,height=0.82\textheight,keepaspectratio]{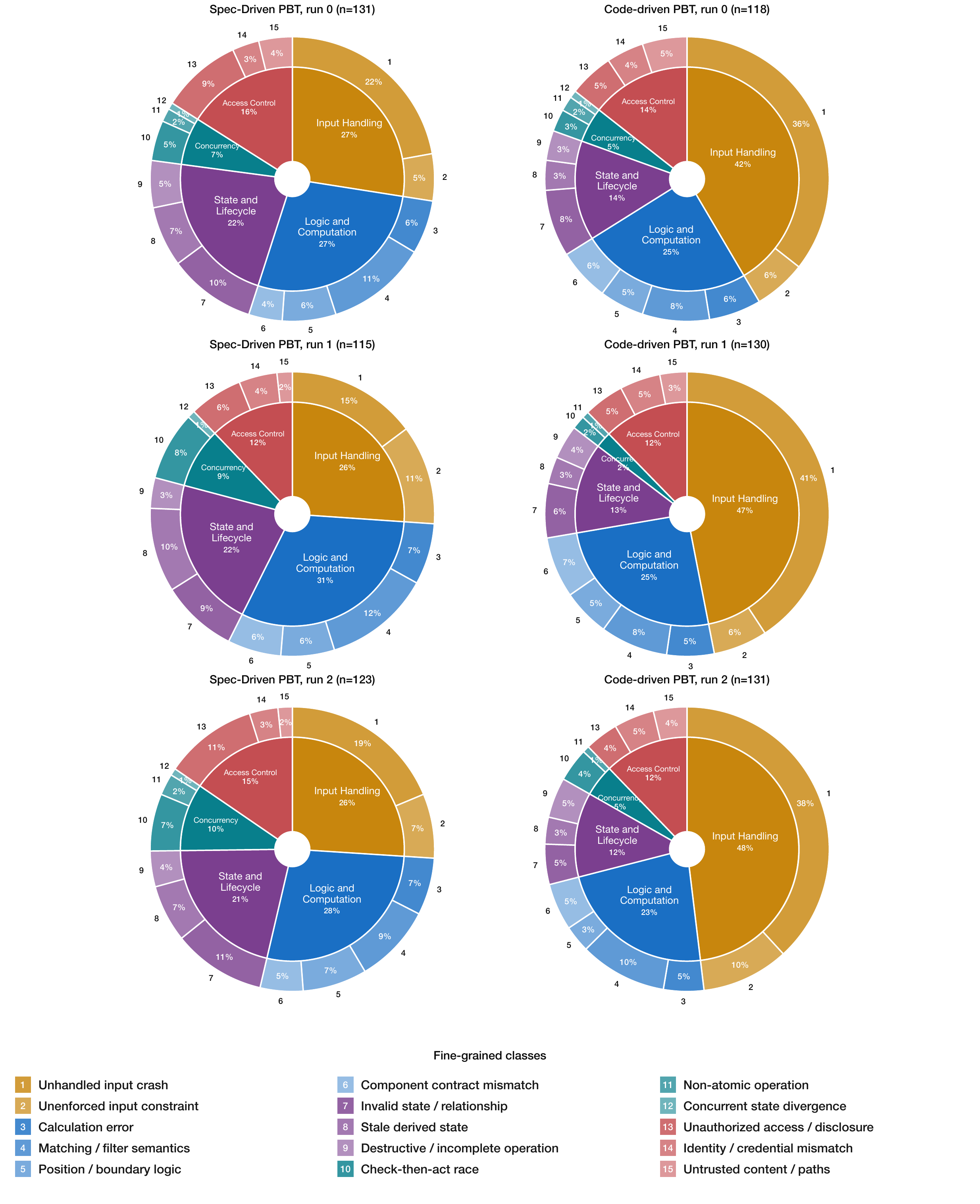}
  \caption{Bug-class composition across three runs per method, using the same sunburst styling and taxonomy as Figure~\ref{fig:bug-taxonomy}. ``Concurrency'' abbreviates ``Concurrency and Atomicity'' inside the smaller plots. Numbered outer slices map to fine classes in the shared legend. A candidate appearing in multiple runs is counted in each of those runs.}
  \label{fig:repeated-runs-taxonomy}
\end{figure}

We examined three runs per method on the same 25 ViBench applications, pooling all six suites per application in one LLM review. Run 0 reuses the submitted suites; its candidate counts differ from Table~\ref{tab:vibench-totals} because they were reviewed in this larger pool. We applied the paper's taxonomy to the 270 distinct candidates. Table~\ref{tab:repeated-runs} gives counts by run and Figure~\ref{fig:repeated-runs-taxonomy} shows their class distributions.

\begin{table}[!ht]
  \caption{Tests, reviewer-attributed candidate bugs, and class counts across three runs per method. Class entries give counts (share of that run's candidate bugs); state/concurrency combines state, lifecycle, data integrity, concurrency, and atomicity.}
  \label{tab:repeated-runs}
  \centering
  \small
  \begin{tabular}{@{}llrrrr@{}}
    \toprule
    Method & Run & Tests & Candidate bugs & Input handling & State/concurrency \\
    \midrule
    Spec-Driven & 0 & 566 & 131 & 36 (27\%) & 38 (29\%) \\
                & 1 & 519 & 115 & 30 (26\%) & 35 (30\%) \\
                & 2 & 538 & 123 & 32 (26\%) & 38 (31\%) \\
    \addlinespace
    Code-driven & 0 & 1384 & 118 & 49 (42\%) & 23 (19\%) \\
                & 1 & 1417 & 130 & 61 (47\%) & 20 (15\%) \\
                & 2 & 1339 & 131 & 63 (48\%) & 22 (17\%) \\
    \bottomrule
  \end{tabular}
\end{table}

\subsection{Qualitative bug taxonomy}
\label{app:bug-taxonomy}

The taxonomy was induced from all 169 canonical ViBench candidate-bug reports.
Each candidate receives exactly one of the fifteen fine-grained labels nested
under five broad categories. The hierarchy and definitions are presented
together below.

Following the ViBench qualitative-analysis protocol~\citep{zhong2026vibench}, we supplied an LLM-driven agent with all candidate bugs and asked it to induce at most five broad and fifteen fine-grained fault classes. A separate Claude Opus 5 agent then applied this taxonomy twice to the method-exclusive bugs, with method labels withheld at both stages. The two classification passes agreed on 98\% of broad and 97\% of fine-grained labels.

\newcommand{\taxdef}[2]{\par\noindent\hangindent=1.5em\hangafter=1
  \hspace*{0.75em}\textbf{#1}\enspace #2\par}

\subsubsection{Taxonomy and definitions}

\paragraph{Input Handling and Robustness.} The application accepts a value from a client, a URL, a file upload or its own configuration and then uses it without checking that it is usable. Either the request dies in a lower layer, or an unacceptable value is taken in and used as if it were valid.
\taxdef{Unhandled input reaches a lower layer.}{A supplied value the code cannot process propagates into the database driver, a parser, arithmetic or an absent API and aborts the operation, instead of being answered with the ordinary client-error the same endpoint already produces for comparable input.}
\taxdef{Declared input constraint not enforced.}{A limit or allowed-value rule that the product states, or that another entry point already applies, is missing, applied on only some paths, or defeated by type coercion, so an out-of-domain value is accepted and used.}

\paragraph{Domain Logic, Computation and Internal Contracts.} The application runs its own rules and produces the wrong answer: a wrong number, a wrong result set, a wrong position in a document or list, or a value that one component writes and another misreads.
\taxdef{Wrong computed or aggregated value.}{A formula, allocation or aggregation is implemented incorrectly, so the number the application reports or stores is not the one its own rules define.}
\taxdef{Wrong search, filter, or lookup result.}{A search, filter or record lookup returns items it should not, or omits items it should return, because the matching rule does not mean what the product says it means.}
\taxdef{Wrong position, boundary, or interval.}{Positions, ranges, addresses or geometry are derived or interpreted incorrectly, so content lands in the wrong place, is cut in the wrong place, or is not reachable at all.}
\taxdef{Component contract mismatch.}{One part of the system writes, names, encodes or advertises something in a way another part does not expect, so the two behave as if they were built against different specifications.}

\paragraph{State, Lifecycle and Data Integrity.} Persisted records end up in a shape the product forbids, keep values that a later change invalidated, or lose data because an operation cleaned up, cascaded or reversed itself incorrectly.
\taxdef{Illegal state, transition, or relationship.}{An operation moves an object into a status, or creates a link between objects, that the product's own rules say cannot exist - usually because the current state or the surrounding relationships were never re-checked.}
\taxdef{Derived state not refreshed.}{Something computed or cached from other data - a result, a flag, a snapshot, a pending record or a displayed indicator - keeps its old value after the data it was derived from changed.}
\taxdef{Destructive or incomplete operation.}{An operation deletes or overwrites more than it should, leaves related records unusable, or records an inverse that cannot faithfully restore what was there before.}

\paragraph{Concurrency and Atomicity.} The defect appears only when two operations overlap, interleave, or when one operation fails part-way through. A single sequential run behaves correctly.
\taxdef{Unsynchronized check-then-write.}{A limit, uniqueness rule or counter is verified by a read and then applied by a separate write, with no lock or constraint tying the two together, so overlapping requests all pass the check.}
\taxdef{Partial effects from a non-atomic operation.}{A multi-step operation is not wrapped as a unit, so a rejection or error part-way through leaves some steps applied and the caller told the operation did not happen.}
\taxdef{Concurrent state divergence.}{Two or more participants acting at the same time end up with different views of shared state, or one participant's update silently absorbs or discards another's, even though every individual request succeeded.}

\paragraph{Access Control, Identity and Untrusted Content.} Content or capability reaches a party that should not have it, an account cannot be identified consistently, or untrusted input is emitted or resolved without being neutralised.
\taxdef{Missing or inconsistent access control.}{An endpoint, field or embedded object is served or mutated without the visibility, ownership or role check that the rest of the application applies to the same data.}
\taxdef{Identity or credential mismatch.}{The value used to identify an account is stored, compared or generated inconsistently, so accounts become unreachable, duplicated, or blocked by a name collision.}
\taxdef{Untrusted content or unsafe path.}{User-supplied text, files or URLs are interpolated into markup, served with an attacker-chosen content type, followed as a redirect target, or joined onto a filesystem path without being neutralised or confined.}

\subsubsection{Results}

\begin{table}[h]
\caption{Distribution of method-exclusive ViBench candidate bugs across the five broad categories.}
\label{tab:bug-taxonomy-results}
\centering
\small
\begin{tabular}{@{}lrr@{}}
\toprule
Broad category & Spec-Driven only ($n=56$) & Code-driven only ($n=43$) \\
\midrule
Input Handling and Robustness & 9 (16.1\%) & 21 (48.8\%) \\
Domain Logic, Computation and Internal Contracts & 15 (26.8\%) & 8 (18.6\%) \\
State, Lifecycle and Data Integrity & 19 (33.9\%) & 5 (11.6\%) \\
Concurrency and Atomicity & 7 (12.5\%) & 4 (9.3\%) \\
Access Control, Identity and Untrusted Content & 6 (10.7\%) & 5 (11.6\%) \\
\bottomrule
\end{tabular}
\end{table}

\subsection{Bug Reports}
\label{app:examples}
The agents' own reports for the five bugs shared as case studies in
Section~\ref{sec:discussion}. Each was written by the agent
at the end of its run.

\subsubsection{Found only by Spec-Driven PBT: a follow of a public account stays pending}
\begin{lstlisting}[basicstyle=\ttfamily\scriptsize,numbers=none,columns=flexible]
# Bug Report: following a public account stays stuck at "pending" when a stale follow request
exists

**Target**: `server/routers/follows.py::follow`
**Severity**: Medium
**Bug Type**: Logic
**Date**: 2026-09-09

## Summary

`follow` returns early on *any* existing `follows` row: `if existing: return {"status":
existing["status"]}`. A pending request created while the target was private therefore
survives the target switching to public, and every later follow attempt keeps answering
`pending`. The follower can never follow the now-public account (the UI shows
"Requested"), and the account's posts never reach their home feed, until the requester
happens to cancel the request first.

## Property-Based Test

```python
PRIORS = ["none", "already_following", "unfollowed", "request_declined_while_private",
          "request_pending_from_private_days", "blocked_then_unblocked"]


@given(prior=st.sampled_from(PRIORS))
@settings(max_examples=len(PRIORS) * 3, deadline=None,
          suppress_health_check=[HealthCheck.too_slow])
def test_following_a_public_account_takes_effect_immediately(prior):
    """PRD: "Following a public account takes effect immediately." """
    target = make_user("public")
    viewer = make_user("public")
    post = target.create_post(caption="hello from a public account")
    _reach_state(viewer, target, prior)
    assert target.get(f"/users/{target.username}").json()["user"]["visibility"] == "public"

    r = viewer.follow(target)
    assert r.status_code == 200, r.text
    assert r.json()["status"] == "accepted", (
        f"prior={prior}: following a public account returned {r.json()['status']!r}"
    )
    rel = viewer.relationship_to(target)
    assert rel["following"] is True and rel["requested"] is False, (prior, rel)
    feed_ids = [p["id"] for p in viewer.get("/feed").json()["posts"]]
    assert post["id"] in feed_ids, f"prior={prior}: followed post missing from the home feed"
```

where the failing prior state is reached by

```python
elif prior == "request_pending_from_private_days":
    target.set_visibility("private")
    viewer.follow(target)
    target.set_visibility("public")
```

**Failing input**: `prior='request_pending_from_private_days'` ---
`AssertionError: following a public account returned 'pending'`.

## Reproducing the Bug

```python
import requests, uuid

API = "http://localhost:8000/api"

def signup(visibility="public"):
    name = "u" + uuid.uuid4().hex[:10]
    r = requests.post(f"{API}/auth/signup", json={
        "username": name, "email": f"{name}@example.com", "display_name": name,
        "password": "password123", "visibility": visibility})
    return name, {"Authorization": "Bearer " + r.json()["token"]}

target, th = signup("private")
viewer, vh = signup()

assert requests.post(f"{API}/users/{target}/follow", headers=vh).json()["status"] == "pending"
requests.patch(f"{API}/me", json={"visibility": "public"}, headers=th)

r = requests.post(f"{API}/users/{target}/follow", headers=vh)
print(r.json())
rel = requests.get(f"{API}/users/{target}", headers=vh).json()["user"]["relationship"]
print(rel)
assert r.json()["status"] == "accepted", "public follow did not take effect"
```

Output:

```
{'status': 'pending'}
{'is_self': False, 'following': False, 'requested': True, ...}
```

## Why This Is A Bug

The PRD states plainly: "Following a **public** account takes effect immediately.
Following a **private** account creates a pending **follow request** the owner must
approve or decline". Pending requests only exist for private accounts, so once the target
is public a follow must produce an accepted edge. Instead the account is unfollowable
through the normal path, its posts are silently missing from the follower's
reverse-chronological home feed (the profile shows them, the feed does not), and the
now-public owner is left holding an approval queue the PRD never describes for a public
account.

## Fix

Only honour an existing row when it is already accepted, or when the target is still
private; otherwise promote the stale request.

```diff
--- a/server/routers/follows.py
+++ b/server/routers/follows.py
@@
     existing = query_one(
                  "SELECT * FROM follows WHERE follower_id = %s AND followee_id = %s",
         (user["id"], target["id"])
     )
-    if existing:
-        return {"status": existing["status"]}
     status = "accepted" if target["visibility"] == "public" else "pending"
+    if existing:
+        if existing["status"] == "accepted" or status == "pending":
+            return {"status": existing["status"]}
+        execute(
+            "UPDATE follows SET status = 'accepted' WHERE follower_id = %s AND followee_id =
%s",
+            (user["id"], target["id"]),
+        )
+        return {"status": "accepted"}
     execute(
                  "INSERT INTO follows (follower_id, followee_id, status) VALUES (%s,%s,%s) ON
         CONFLICT DO NOTHING",
         (user["id"], target["id"], status),
     )
     return {"status": status}
```

(Converting outstanding requests at the moment the account switches to public, in
`users.update_profile`, would additionally clear the stale approval queue.)
\end{lstlisting}
\subsubsection{Found only by Spec-Driven PBT: accepted commits are silently lost}
\begin{lstlisting}[basicstyle=\ttfamily\scriptsize,numbers=none,columns=flexible]
# Bug Report: backend.routes_git.make_commit Loses Commits and Silently Discards Merges (Missing
Branch Lock)

**Target**: `backend/routes_git.py::make_commit` (`POST /api/repos/{owner}/{name}/commits`)
**Severity**: High
**Bug Type**: Logic
**Date**: 2026-09-09

## Summary

`make_commit` reads `branch.head_id`, builds the new tree from it, and writes
`branch.head_id` back without taking a row lock on the branch --- unlike
`cherry_pick`, `revert` and `merge_pull`, which all use
`select(Branch)...with_for_update()`. Concurrent writes to one branch therefore
lost-update each other: every request returns `201 Created` but only the last
writer's commit stays on the branch. When a plain commit races a pull-request
merge on the same branch, the merge is the one that gets discarded: the PR is
left in state `merged` with a `merge_commit_id` that is not in the base
branch's history, the head branch's changes never reach the base, and the PR
cannot be reopened.

## Property-Based Test

```python
@CONC
@given(payload=st.text(alphabet="abcdef", min_size=1, max_size=6))
def test_merged_pull_request_always_reaches_its_base_branch(make_user, payload):
    owner = make_user()
    reviewer = make_user()
    repo = owner.new_repo()
    owner.post(
        f"/api/repos/{repo}/collaborators",
        {"username": reviewer.username, "role": "write"},
    )
    owner.commit(
        repo, "main", "init", [{"path": "base.txt", "action": "create", "content": "1\n"}]
    )
    owner.post(f"/api/repos/{repo}/branches", {"name": "feature", "source": "main"})
    owner.commit(
        repo,
        "feature",
        "feature work",
        [{"path": "feature.txt", "action": "create", "content": payload + "\n"}],
    )
    pr = owner.post(
        f"/api/repos/{repo}/pulls", {"title": "p", "head": "feature", "base": "main"}
    )
    number = pr["number"]
    reviewer.post(f"/api/repos/{repo}/pulls/{number}/reviews", {"state": "approve"})

    def do_merge():
        c = _sibling(owner)
        try:
            return c.req("POST", f"/api/repos/{repo}/pulls/{number}/merge")
        finally:
            c.close()

    def do_commit():
        c = _sibling(owner)
        try:
            return c.req(
                "POST",
                f"/api/repos/{repo}/commits",
                json={
                    "branch": "main",
                    "message": "unrelated direct commit",
                    "changes": [
                        {"path": "direct.txt", "action": "create", "content": "d\n"}
                    ],
                },
            )
        finally:
            c.close()

    merge_res, _commit_res = _run_parallel([do_merge, do_commit])

    pr_state = owner.get(f"/api/repos/{repo}/pulls/{number}")
    history = [
        c["id"]
        for c in owner.get(f"/api/repos/{repo}/commits", params={"ref": "main"})["commits"]
    ]
    tree = owner.tree(repo, "main")

    if merge_res.status_code < 400:
        assert pr_state["state"] == "merged"
        assert pr_state["merge_commit_id"] in history
        assert "feature.txt" in tree
    else:
        assert pr_state["state"] == "open", pr_state["state"]
```

**Failing input**: `payload='a'` --- any value fails; the minimal shrink of the
commit-only variant is `n_writers=2`.

## Reproducing the Bug

```python
import concurrent.futures as cf

import httpx

BASE = "http://localhost:8000"


def signup(username):
    c = httpx.Client(base_url=BASE, timeout=60)
    c.post(
        "/api/auth/signup",
        json={
            "name": username,
            "username": username,
            "email": f"{username}@example.com",
            "password": "pw123456",
        },
    )
    return c


owner = signup("race-owner")
reviewer = signup("race-reviewer")
repo = "race-owner/race-repo"

owner.post("/api/repos", json={"name": "race-repo", "visibility": "public"})
owner.post(
    f"/api/repos/{repo}/collaborators", json={"username": "race-reviewer", "role": "write"}
)
owner.post(
    f"/api/repos/{repo}/commits",
    json={
        "branch": "main",
        "message": "init",
        "changes": [{"path": "base.txt", "action": "create", "content": "1\n"}],
    },
)
owner.post(f"/api/repos/{repo}/branches", json={"name": "feature", "source": "main"})
owner.post(
    f"/api/repos/{repo}/commits",
    json={
        "branch": "feature",
        "message": "feature work",
        "changes": [{"path": "feature.txt", "action": "create", "content": "f\n"}],
    },
)
n = owner.post(
    f"/api/repos/{repo}/pulls", json={"title": "p", "head": "feature", "base": "main"}
).json()["number"]
reviewer.post(f"/api/repos/{repo}/pulls/{n}/reviews", json={"state": "approve"})


def merge():
    c = httpx.Client(base_url=BASE, cookies=owner.cookies, timeout=60)
    return c.post(f"/api/repos/{repo}/pulls/{n}/merge")


def commit():
    c = httpx.Client(base_url=BASE, cookies=owner.cookies, timeout=60)
    return c.post(
        f"/api/repos/{repo}/commits",
        json={
            "branch": "main",
            "message": "direct",
            "changes": [{"path": "direct.txt", "action": "create", "content": "d\n"}],
        },
    )


with cf.ThreadPoolExecutor(max_workers=2) as pool:
    f1, f2 = pool.submit(merge), pool.submit(commit)
    r_merge, r_commit = f1.result(), f2.result()

pr = owner.get(f"/api/repos/{repo}/pulls/{n}").json()
history = [
        c["id"] for c in owner.get(f"/api/repos/{repo}/commits", params={"ref":
    "main"}).json()["commits"]
]
paths = owner.get(f"/api/repos/{repo}/tree", params={"ref": "main"}).json()["all_paths"]

print("merge:", r_merge.status_code, "commit:", r_commit.status_code)
print("pr state:", pr["state"], "merge commit in history:", pr["merge_commit_id"] in history)
print("main tree:", paths)
print("reopen:", owner.post(f"/api/repos/{repo}/pulls/{n}/reopen").status_code)
```

Output:

```
merge: 200 commit: 201
pr state: merged merge commit in history: False
main tree: ['base.txt', 'direct.txt']
reopen: 400
```

## Why This Is A Bug

The PRD states that "A commit advances the branch's tip and updates its tree
accordingly", that "History is append-only", and --- for merges specifically ---
that "Merge-gate evaluation and the merge's application are atomic against the
base as it stands at that moment: when two merges race for the same base, the
first to commit takes effect, and any competing merge is re-evaluated against
the advanced base."

Here neither holds. Both requests read the same `head_id`, each creates a
commit parented on it, and the second `UPDATE branches SET head_id` overwrites
the first. The overwritten commit is orphaned: it exists as a row but is not
reachable from any branch, so its changes vanish while the API reported
success. In the merge case the result is worse than a lost commit --- the PR's
recorded state (`merged`, plus a `merge_commit_id`) contradicts the repository,
`feature.txt` is absent from `main`, and "A merged pull request cannot be
reopened" makes the state unrecoverable through the API.

That `cherry_pick`, `revert` and `merge_pull` all lock the branch row while
`make_commit` does not shows this is an omission rather than a design choice;
the merge/merge race is in fact handled correctly (verified separately: one
merge wins, the other gets 409).

## Fix

Lock the branch row before reading its tip, matching the other write paths.

```diff
--- a/backend/routes_git.py
+++ b/backend/routes_git.py
@@ -277,9 +277,14 @@ def make_commit(
 ):
     repo = get_repo(db, owner, name, user)
     require_write(db, repo, user)
-    branch = get_branch(db, repo.id, payload.branch)
+    branch = db.execute(
+        select(Branch)
+        .where(Branch.repo_id == repo.id, Branch.name == payload.branch)
+        .with_for_update()
+    ).scalar_one_or_none()
     if branch is None:
         raise HTTPException(status_code=404, detail="Branch not found")
+    db.refresh(branch)
     message = (payload.message or "").strip()
     if not message:
         raise HTTPException(status_code=400, detail="A commit message is required")
```
\end{lstlisting}
\subsubsection{Found only by Code-driven PBT: a malformed quantity crashes the cart}
\begin{lstlisting}[basicstyle=\ttfamily\scriptsize,numbers=none,columns=flexible]
# Bug Report: Unparseable numeric fields return HTTP 500 instead of the documented 4xx error
codes

**Target**: `routes_seller._validate_listing`, `routes_shop.add_to_cart`,
`mutations.place_order`, `auth.validate_name`
**Severity**: Medium
**Bug Type**: Crash
**Date**: 2026-09-04

## Summary

The write routes validate their numeric inputs inconsistently: `stock`, `rating` and
`dealQuantity` are wrapped in `try/except (TypeError, ValueError)` and answer with a
`400` and a stable error code, while the adjacent `price` / `priceCents` / `qty` /
`expectedTotalCents` fields call `int()` or `money.to_cents()` bare. A blank or
non-numeric value there escapes as an unhandled exception and is turned into a
`500 SERVER_ERROR` by `main.unhandled_handler`, whose response also echoes the raw
exception text (`"message": str(exc)`).

## Property-Based Test

```python
bad_money = st.sampled_from(["", "abc", "$19.99", "19,99", "nan", "1.2.3", " "])


@given(price=bad_money)
def test_unparseable_listing_price_is_a_client_error(api, seller, price):
    status, body = api.call(
        "POST", "/api/seller/listings",
        {"title": uniq("t-"), "category": "General", "price": price, "stock": 1},
        seller["token"],
    )
    assert status == 400, body
    assert body["code"] == "INVALID_PRICE"
```

**Failing input**: `price=""` (an empty price field) -> `500`,
`{"code": "SERVER_ERROR", "message": "[<class 'decimal.ConversionSyntax'>]"}`.

The sibling test `test_unparseable_listing_stock_is_a_client_error` passes with the
same kind of input, which is what pins the intended contract.

## Reproducing the Bug

```python
import json, urllib.error, urllib.request, uuid

BASE = "http://127.0.0.1:8000"


def call(method, path, body=None, token=None):
    req = urllib.request.Request(BASE + path, data=json.dumps(body).encode(), method=method)
    req.add_header("content-type", "application/json")
    if token:
        req.add_header("authorization", "Bearer " + token)
    try:
        with urllib.request.urlopen(req) as resp:
            return resp.status, json.loads(resp.read())
    except urllib.error.HTTPError as exc:
        return exc.code, json.loads(exc.read())


tag = uuid.uuid4().hex[:8]
_, seller = call("POST", "/api/auth/signup", {
    "name": "s", "email": f"s{tag}@example.test", "password": "hunter2pw",
    "role": "seller", "storeName": f"store-{tag}"})
token = seller["token"]
_, shopper = call("POST", "/api/auth/signup", {
    "name": "b", "email": f"b{tag}@example.test", "password": "hunter2pw", "role": "shopper"})

print(call("POST", "/api/seller/listings",
           {"title": "t", "category": "General", "price": "", "stock": 1}, token)[0])
print(call("POST", "/api/seller/listings",
           {"title": "t", "category": "General", "priceCents": "1.5", "stock": 1}, token)[0])
print(call("POST", "/api/seller/listings",
           {"title": "t", "category": "General", "priceCents": 500, "stock": "abc"}, token)[0])
print(call("POST", "/api/auth/signup",
           {"name": 123, "email": f"n{tag}@example.test",
            "password": "hunter2pw", "role": "shopper"})[0])
```

```
500      <- POST /api/seller/listings   price: ""
500      <- POST /api/seller/listings   priceCents: "1.5"
400      <- POST /api/seller/listings   stock: "abc"   (handled, for contrast)
500      <- POST /api/auth/signup       name: 123
```

Also reproducing as `500`:

| Request | Field | Code path |
| --- | --- | --- |
| `POST /api/cart` | `qty: "1.5"` | `routes_shop.add_to_cart` -> `int(...)` |
| `PATCH /api/cart/{id}` | `qty: "abc"` | `routes_shop.update_cart_item` -> `int(...)` |
| `POST /api/orders` | `expectedTotalCents: "abc"` | `mutations.place_order` -> `int(...)` |
| `POST /api/auth/signup` | `name: 123` | `auth.validate_name` -> `(name or "").strip()` |

## Why This Is A Bug

* `services.err` and the `HTTPException` handler define the API contract: a
  `{code, message}` body with a 4xx status. `_validate_listing` already produces
  `INVALID_PRICE` for a non-positive price and `INVALID_STOCK` for an unparseable
  stock --- the price branch simply lacks the same guard, three lines above the one
  that has it.
* `auth.validate_password` explicitly rejects non-string input
  (`if password is None or not isinstance(password, str)`), so type-safety on
  `validate_name` is clearly intended too.
* An empty string is what an HTML form submits for a field the seller left blank, so
  this is ordinary user input, not a synthetic edge case: the seller sees "Unexpected
  server error" instead of "Enter a price greater than $0.00."
* The 500 body also leaks internals (`str(exc)` includes the failing SQL statement or
  `decimal.ConversionSyntax`).

## Fix

Give the numeric fields the same guard the neighbouring ones already have.

```diff
--- a/backend/routes_seller.py
+++ b/backend/routes_seller.py
@@ def _validate_listing(payload: dict, partial: bool = False) -> dict:
     if "price" in payload or "priceCents" in payload or not partial:
-        cents = int(payload["priceCents"]) if payload.get("priceCents") is not None else
to_cents(payload.get("price"))
+        try:
+            cents = (
+                int(payload["priceCents"])
+                if payload.get("priceCents") is not None
+                else to_cents(payload.get("price"))
+            )
+        except (TypeError, ValueError, ArithmeticError):
+            raise err(400, "INVALID_PRICE", "Enter a price greater than $0.00.")
         if cents <= 0:
             raise err(400, "INVALID_PRICE", "Enter a price greater than $0.00.")
         out["price_cents"] = cents

--- a/backend/routes_shop.py
+++ b/backend/routes_shop.py
@@ def add_to_cart(payload: dict = Body(...), db: Session = Depends(get_db), user: User =
Depends(require_shopper)):
-    qty = int(payload.get("qty") or 1)
+    try:
+        qty = int(payload.get("qty") or 1)
+    except (TypeError, ValueError):
+        raise err(400, "INVALID_QTY", "Quantity must be at least 1.")
     if qty <= 0:
         raise err(400, "INVALID_QTY", "Quantity must be at least 1.")

--- a/backend/auth.py
+++ b/backend/auth.py
@@ def validate_name(name: str) -> str:
-    n = (name or "").strip()
+    if name is not None and not isinstance(name, str):
+        raise HTTPException(
+            status_code=400,
+            detail={"code": "INVALID_NAME", "message": "Name must be 1-80 characters and cannot
be only whitespace."},
+        )
+    n = (name or "").strip()
```

The same treatment is needed for `update_cart_item`'s `qty`, `place_order`'s
`expectedTotalCents`, and the variant/deal price parsing in `routes_seller`
(`create_listing`, `update_listing`, `create_deal`).
\end{lstlisting}
\subsubsection{Found only by Code-driven PBT: a username is stored under a different string}
\begin{lstlisting}[basicstyle=\ttfamily\scriptsize,numbers=none,columns=flexible]
# Bug Report: `core.views.register_view` stores a different username than the one validated
(NFKC)

**Target**: `core.views.register_view` / `core.forms.RegistrationForm.clean_username`
**Severity**: Medium
**Bug Type**: Logic (with a Crash symptom)
**Date**: 2026-09-09

## Summary

`RegistrationForm.clean_username` checks availability of the *raw* username, but
`User.objects.create_user()` stores the NFKC-normalised form of it. For any
username that is not NFKC-stable (e.g. `'<U+00AA>'`, `'<U+FB00>'`, full-width `'<U+FF21>'`) the
account is created under a different name, so the member can never log in
again; and if the normalised name is already taken the uniqueness check is
bypassed and the view dies with an `IntegrityError` (HTTP 500).

## Property-Based Test

```python
@given(username=unicode_usernames, email=emails, password=passwords)
def test_credentials_used_to_register_can_be_used_to_log_in(self, username, email, password):
    code = make_invite_code()
    self.client.post('/register/', registration_payload(username, email, password, code.code))
    self.client.get('/logout/')

    response = self.client.post('/login/', {'username': username, 'password': password})
    self.assertEqual(response.status_code, 302,
                     f'cannot log in as {username!r} right after registering with it')


@given(username=unicode_usernames, password=passwords)
def test_second_registration_with_same_username_is_rejected(self, username, password):
    """"Username already taken." -- the same username never registers twice."""
    code = make_invite_code()
    first = self.client.post(
        '/register/', registration_payload(username, 'first@example.com', password, code.code))
    assume(first.status_code == 302)
    self.client.get('/logout/')

    second = self.client.post(
        '/register/', registration_payload(username, 'second@example.com', password, code.code))
    self.assertEqual(second.status_code, 200, 'duplicate username was not rejected cleanly')
    self.assertEqual(User.objects.count(), 1)
```

**Failing input**: `username='<U+00AA>'` (NFKC-normalises to `'a'`)

## Reproducing the Bug

```python
import os

os.environ['DJANGO_SETTINGS_MODULE'] = 'creative_community.settings'
os.environ['POSTGRES_DATABASE_URL'] =
'postgres://postgres:postgres@127.0.0.1:5432/creative_community'

import django

django.setup()

from django.contrib.auth.models import User
from django.test import Client

from core.models import InviteCode

InviteCode.objects.get_or_create(
    code='REPRO', defaults={'code_type': 'multi', 'max_uses': 100}
)

client = Client()
client.post('/register/', {
    'username': '<U+00AA>', 'email': 'one@example.com',
    'password': 'pw', 'password_confirm': 'pw', 'invite_code': 'REPRO',
})
print(list(User.objects.values_list('username', flat=True)))   # ['a']  -- not '<U+00AA>'

client.get('/logout/')
print(client.post('/login/', {'username': '<U+00AA>', 'password': 'pw'}).status_code)  # 200 =
login failed

client.post('/register/', {
    'username': '<U+00AA>', 'email': 'two@example.com',
    'password': 'pw', 'password_confirm': 'pw', 'invite_code': 'REPRO',
})   # django.db.utils.IntegrityError: duplicate key value ... (username)=(a)
```

## Why This Is A Bug

`clean_username` promises "Username already taken." for names that are in use,
and the account the view creates is expected to be the one the visitor typed:
`register_view` logs them in immediately, so the *only* later route into the
account is `login_view`, which looks the username up verbatim
(`ModelBackend` -> `get_by_natural_key`). Two concrete consequences:

1. **Silent lockout** -- registering as `'<U+00AA>'` creates the user `'a'`; the invite
   code is consumed and the member is locked out as soon as the session ends.
2. **Unhandled 500** -- when the normalised name already exists, the uniqueness
   check misses it and the `create_user()` call violates the
   `auth_user_username_key` constraint. `MemberProfile`/`InviteCodeUsage` are not
   created but `InviteCode.current_uses` has not been touched yet, so the state
   is merely wasted, not corrupted.

Django's own `UserCreationForm` avoids this by normalising the username in
`clean_username`; this hand-written form does not.

## Fix

Normalise the username in the form so that validation and storage agree:

```diff
--- a/core/forms.py
+++ b/core/forms.py
@@
 from django import forms
 from django.contrib.auth.models import User
 from .models import MemberProfile, InviteCode, Event
@@
     def clean_username(self):
-        username = self.cleaned_data['username']
+        username = User.normalize_username(self.cleaned_data['username'])
+        if len(username) > 150:
+            raise forms.ValidationError('Username is too long.')
         if User.objects.filter(username=username).exists():
             raise forms.ValidationError('Username already taken.')
         return username
```
\end{lstlisting}
\subsubsection{Found by both methods: the kick rule is written twice and the copies disagree}
\lstinputlisting[basicstyle=\ttfamily\scriptsize,numbers=none,columns=flexible]{generated/reports/discord-bug-005.md}

\newpage
\section*{NeurIPS Paper Checklist}

\begin{enumerate}

\item {\bf Claims}
    \item[] Question: Do the main claims made in the abstract and introduction accurately reflect the paper's contributions and scope?
    \item[] Answer: \answerYes{} 
    \item[] Justification: The abstract and the contribution bullets in Section 1 state the headline results --- $2.4\times$ fewer tests and 169 distinct bugs of which nearly 60\% are found by only one method --- together with the qualitative claim about which bug classes each method surfaces. These are supported by Table~\ref{tab:vibench-totals} and Figure~\ref{fig:bug-taxonomy}, and the scope is confined to the two testbeds described in Section~\ref{sec:setup}.

\item {\bf Limitations}
    \item[] Question: Does the paper discuss the limitations of the work performed by the authors?
    \item[] Answer: \answerYes{} 
    \item[] Justification: Section~\ref{sec:limitations} discusses the absence of ground-truth bug labels on ViBench, the limits of validation on a subset, the incompleteness of property-based testing, and dependence on specification quality. Appendices~\ref{app:invariant-source-comparison} and~\ref{app:repeated-runs} report the controlled comparison and repeated-run results.

\item {\bf Theory assumptions and proofs}
    \item[] Question: For each theoretical result, does the paper provide the full set of assumptions and a complete (and correct) proof?
    \item[] Answer: \answerNA{} 
    \item[] Justification: The paper presents no theoretical results.

    \item {\bf Experimental result reproducibility}
    \item[] Question: Does the paper fully disclose all the information needed to reproduce the main experimental results of the paper to the extent that it affects the main claims and/or conclusions of the paper (regardless of whether the code and data are provided or not)?
    \item[] Answer: \answerYes{} 
    \item[] Justification: Section~\ref{sec:setup} describes both testbeds, the agents' inputs, the model, harness, tools, and review procedure. Appendices~\ref{app:prompt} and~\ref{app:code-driven-prompt} provide the full agent prompts; Appendix~\ref{app:detail} reports per-application ViBench results. Appendices~\ref{app:invariant-source-comparison} and~\ref{app:repeated-runs} describe the additional comparisons.

\item {\bf Open access to data and code}
    \item[] Question: Does the paper provide open access to the data and code, with sufficient instructions to faithfully reproduce the main experimental results, as described in supplemental material?
    \item[] Answer: \answerYes{} 
    \item[] Justification: Code and data are provided as supplementary material. Independently of that, Appendix~\ref{app:prompt} reproduces the full prompt verbatim, which is the whole of our method, and both testbeds are publicly available.

\item {\bf Experimental setting/details}
    \item[] Question: Does the paper specify all the training and test details (e.g., data splits, hyperparameters, how they were chosen, type of optimizer) necessary to understand the results?
    \item[] Answer: \answerYes{} 
    \item[] Justification: Section~\ref{sec:setup} specifies the agents, reviewer, and taxonomy classifier. Appendices~\ref{app:prompt} and~\ref{app:code-driven-prompt} give the agent prompts, Appendix~\ref{app:detail} the per-application breakdown and execution environment, and Appendices~\ref{app:invariant-source-comparison} and~\ref{app:repeated-runs} the additional experimental designs.

\item {\bf Experiment statistical significance}
    \item[] Question: Does the paper report error bars suitably and correctly defined or other appropriate information about the statistical significance of the experiments?
    \item[] Answer: \answerYes{} 
    \item[] Justification: Appendix~\ref{app:efficiency} reports a paired two-sided sign test over the 25 ViBench applications, treating each application as one paired observation: Spec-Driven PBT writes fewer tests on all 25 and reaches a higher bugs-per-test ratio on all 25, $p \approx 6 \times 10^{-8}$ in both cases. Appendix~\ref{app:repeated-runs} reports results from three runs per method.

\item {\bf Experiments compute resources}
    \item[] Question: For each experiment, does the paper provide sufficient information on the computer resources (type of compute workers, memory, time of execution) needed to reproduce the experiments?
    \item[] Answer: \answerYes{} 
    \item[] Justification: Appendix~\ref{app:cost} gives the executor configuration --- one Modal sandbox per generation cell, a Linux/amd64 Docker container with 2 cores and 4 GiB of memory running Ubuntu 24.04, no accelerators --- together with agent wall-clock time, token counts and API cost, reported per cell and in total for each method.
    
\item {\bf Code of ethics}
    \item[] Question: Does the research conducted in the paper conform, in every respect, with the NeurIPS Code of Ethics \url{https://neurips.cc/public/EthicsGuidelines}?
    \item[] Answer: \answerYes{} 
    \item[] Justification: The research conforms with the NeurIPS Code of Ethics in every respect.

\item {\bf Broader impacts}
    \item[] Question: Does the paper discuss both potential positive societal impacts and negative societal impacts of the work performed?
    \item[] Answer: \answerYes{} 
    \item[] Justification: The Conclusion states the positive impact: the method contributes to the verifiability of agentic coding by surfacing, in the specification's own vocabulary, the requirements an implementation breaks. Section~\ref{sec:limitations} names the corresponding risk: property-based testing is not exhaustive, so a suite that passes is not evidence of correctness, and nothing in this work argues that human review can be dispensed with. We are aware of no other negative societal impact specific to the method; it produces failing tests and bug reports for code the user already controls, and the defects discussed here are in benchmark applications rather than deployed systems.
    
\item {\bf Safeguards}
    \item[] Question: Does the paper describe safeguards that have been put in place for responsible release of data or models that have a high risk for misuse (e.g., pre-trained language models, image generators, or scraped datasets)?
    \item[] Answer: \answerNA{} 
    \item[] Justification: The paper releases no pretrained model, scraped dataset or generative artefact carrying a risk of misuse. The bugs discussed are in benchmark applications, not in deployed software, so no disclosure process applies.

\item {\bf Licenses for existing assets}
    \item[] Question: Are the creators or original owners of assets (e.g., code, data, models), used in the paper, properly credited and are the license and terms of use explicitly mentioned and properly respected?
    \item[] Answer: \answerYes{} 
    \item[] Justification: ViBench~\citep{zhong2026vibench}, PBTBench~\citep{pbtbench}, Hypothesis~\citep{maciver2019hypothesis} and the Agentic PBT baseline~\citep{maaz2025agentic} are each cited where used, and are used within their terms.

\item {\bf New assets}
    \item[] Question: Are new assets introduced in the paper well documented and is the documentation provided alongside the assets?
    \item[] Answer: \answerYes{} 
    \item[] Justification: The prompt that constitutes our method is documented in full in Appendix~\ref{app:prompt}. The code, the generated test suites and the run artifacts accompany the submission as supplementary material.

\item {\bf Crowdsourcing and research with human subjects}
    \item[] Question: For crowdsourcing experiments and research with human subjects, does the paper include the full text of instructions given to participants and screenshots, if applicable, as well as details about compensation (if any)? 
    \item[] Answer: \answerNA{} 
    \item[] Justification: The paper involves no crowdsourcing and no research with human subjects.

\item {\bf Institutional review board (IRB) approvals or equivalent for research with human subjects}
    \item[] Question: Does the paper describe potential risks incurred by study participants, whether such risks were disclosed to the subjects, and whether Institutional Review Board (IRB) approvals (or an equivalent approval/review based on the requirements of your country or institution) were obtained?
    \item[] Answer: \answerNA{} 
    \item[] Justification: The paper involves no research with human subjects, so no review was required.

\item {\bf Declaration of LLM usage}
    \item[] Question: Does the paper describe the usage of LLMs if it is an important, original, or non-standard component of the core methods in this research? Note that if the LLM is used only for writing, editing, or formatting purposes and does \emph{not} impact the core methodology, scientific rigor, or originality of the research, declaration is not required.
    \item[] Answer: \answerYes{} 
    \item[] Justification: LLMs are the object of study and the core of the method, not an incidental tool. Sections~\ref{sec:method} and~\ref{sec:setup} describe the agents, the LLM reviewer, and the taxonomy classifier, which use Claude Opus 5 through the Claude Code harness. Appendices~\ref{app:prompt} and~\ref{app:code-driven-prompt} provide the agents' full prompts.

\end{enumerate}

\end{document}